\documentclass[journal]{IEEEtran}
\usepackage{epsfig,graphics,epsf,color,amsmath,balance,cite,amsfonts,multirow,stfloats,algorithm,algorithmic,mathrsfs,amssymb,color,subfigure,makecell}
\usepackage[dvipdfm,unicode,colorlinks, linkcolor=black,anchorcolor=black,citecolor=black]{hyperref}
\newtheorem{theorem}{Theorem}

\newtheorem{proposition}{Proposition}
\newtheorem{remark}{Remark}

\begin{document}

\title{{Full-Duplex Communication for ISAC: \\Joint Beamforming and Power Optimization}}

\author{Zhenyao He,~\IEEEmembership{Graduate Student Member,~IEEE,}~Wei Xu,~\IEEEmembership{Senior Member,~IEEE,}~Hong Shen,~\IEEEmembership{Member,~IEEE,} Derrick Wing Kwan Ng,~\IEEEmembership{Fellow,~IEEE,} Yonina C. Eldar,~\IEEEmembership{Fellow,~IEEE,} and Xiaohu You,~\IEEEmembership{Fellow,~IEEE}

\thanks{
Part of this work will be presented at 2023 IEEE International Conference on Acoustics, Speech and Signal Processing (ICASSP) \cite{ICASSP}. 

Zhenyao He, Wei Xu, Hong Shen, and Xiaohu You are with the National Mobile Communications Research Laboratory, Southeast University, Nanjing 210096, China (e-mail: \{hezhenyao, wxu, shhseu, xhyu\}@seu.edu.cn).

Derrick Wing Kwan Ng is with the School of Electrical Engineering and Telecommunications, University of New South Wales, Sydney, NSW 2052, Australia (e-mail: w.k.ng@unsw.edu.au).

Yonina C. Eldar is with the Faculty of Mathematics and Computer Science, Weizmann Institute of Science, Rehovot 7610001, Israel (e-mail: yonina@weizmann.ac.il).
}
}

\maketitle

\begin{abstract}
Beamforming design has been widely investigated for integrated sensing and communication (ISAC) systems with full-duplex (FD) sensing and half-duplex (HD) communication, where the base station (BS) transmits and receives radar sensing signals simultaneously while the integrated communication operates in either downlink or uplink. To achieve higher spectral efficiency, in this paper, we extend existing ISAC beamforming design to a general case by considering the FD capability for both radar and communication. Specifically, we consider an FD ISAC system, where the BS performs target detection and communicates with multiple downlink users and uplink users reusing the same time and frequency resources. We jointly optimize the downlink dual-functional transmit signal and the uplink receive beamformers at the BS and the transmit power at the uplink users.
The problems are formulated under two criteria: power consumption minimization and sum rate maximization. The downlink and uplink transmissions are tightly coupled due to both the desired target echo and the undesired interference received at the BS, making the problems challenging.
To handle these issues in both cases, we first determine the optimal receive beamformers in closed forms with respect to the BS transmit beamforming and the user transmit power.
Subsequently, we invoke these results to obtain equivalent optimization problems and propose iterative algorithms to solve them.
In addition, we consider a special case under the power minimization criterion and propose an alternative low complexity design.
Numerical results demonstrate that the optimized FD communication-based ISAC brings tremendous improvements in terms of both power efficiency and spectral efficiency compared to the conventional ISAC with HD communication.
\end{abstract}

\begin{IEEEkeywords}
Integrated sensing and communication (ISAC), full-duplex (FD) communication, joint transceiver optimization, beamforming design.
\end{IEEEkeywords}

\section{Introduction}
Emerging applications, such as Internet-of-Vehicles (IoV) and smart factory, in future wireless networks call for a
demand for reliable sensing and efficient communication to various wireless terminals \cite{W.XuJSTSP2022,W.ShiWCM2022}.
In addition, continuous and aggressive use of frequency spectrum, e.g., millimeter-wave (mmWave), in wireless communications results in overlapped spectrum with conventional radar systems.
These motivate the development of frameworks for sensing-communication integration.
In particular, integrated sensing and communication (ISAC), also known as dual-functional radar-communication and joint radar-communication, has become an appealing technique to address the aforementioned issues and attracted considerable research interest. It has been shown in the literature \cite{overviewJ.Zhang,J.Zhang2022CSTEnabling,overviewF.Liu} that ISAC significantly enhances the spectral efficiency and reduces implemental cost by sharing spectral resources and reusing expensive hardware architectures.

Effective transmit beamforming design is a key to unlock the potential in both multiple-input multiple-output (MIMO) communication systems and MIMO radar systems \cite{P.StoicaMIMOradar1,P.StoicaTSP2007}.
Motivated by this, many works have studied transmit design in multi-antenna ISAC systems by focusing on joint beamforming optimization \cite{F.LiuTWC2018,XLiuTSP2020,HHuaArxiv,Z.LyuArxiv,Z.HeWCL2022}.
Specifically, for conventional MIMO radar systems, a commonly adopted strategy of probing signal design is to manipulate the transmit beampattern through optimizing the covariance matrix of the transmit signal, aiming to maximize the spatial power steered towards desired directions or to minimize the matching error between the transmit signal and a dedicated beampattern \cite{P.StoicaMIMOradar1,P.StoicaTSP2007}.
Leveraging this strategy, the authors in \cite{F.LiuTWC2018} advocated the reuse of transmit signal for both multi-user communication and radar sensing in ISAC systems. Specifically, the beamforming was optimized by minimizing the beampattern matching error, taking into account individual signal-to-interference-plus-noise ratio (SINR) requirements of communication users.
As an alternative, studies \cite{XLiuTSP2020} and \cite{HHuaArxiv} considered similar problems while introducing a dedicated radar signal to facilitate the downlink ISAC. They introduced extra degrees-of-freedom (DoF) to the transmitted signal deliberately to achieve enhanced sensing accuracy.
On the other hand, the authors of \cite{HHuaArxiv} investigated the problem of maximizing the transmit beampattern gain towards the sensing directions in ISAC, while guaranteeing the minimum required SINR of communication users. By imposing the constraint of transmit beampattern gain for sensing, the problems of communication spectral efficiency maximization \cite{Z.LyuArxiv} and energy efficiency maximization \cite{Z.HeWCL2022} were addressed for ISAC.
Note that these works only design the transmit beamforming while the reception of radar echo is not considered.

\begin{figure}[t]
\centering
\subfigure[Integrated sensing with downlink communication.]
{
\begin{minipage}[t]{0.8\linewidth}
\centering
\includegraphics[width=2.5in]{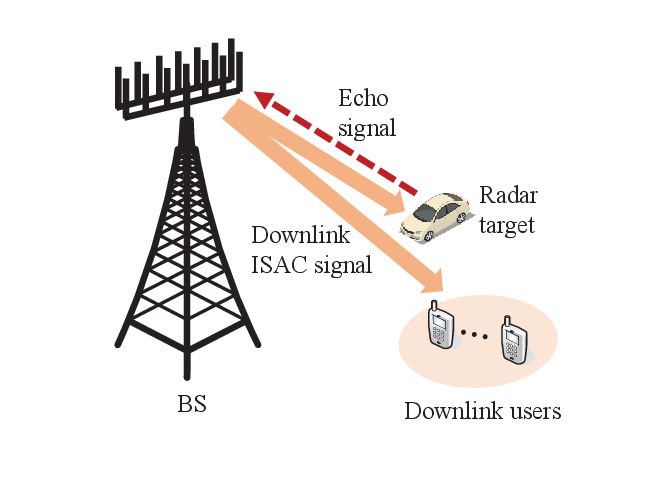}
\end{minipage}
}
\centering
\subfigure[Integrated sensing with uplink communication.]
{
\begin{minipage}[t]{0.8\linewidth}
\centering
\includegraphics[width=2.5in]{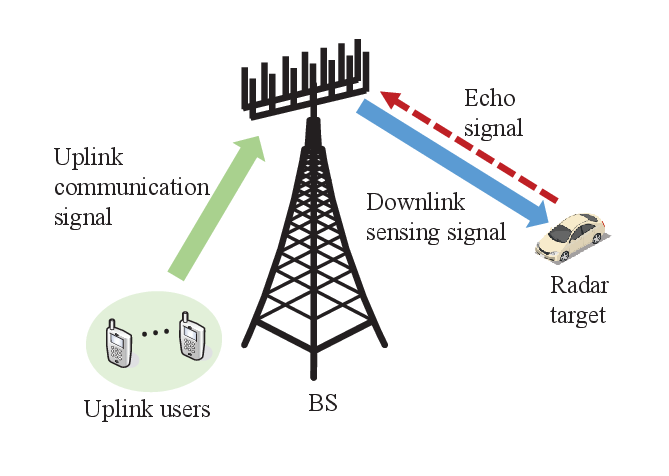}
\end{minipage}
}
\caption{Two cases of ISAC considered in \cite{Liu2022CRB,C.Tsinos2021JSTSP,J.Pritzker2022Arxiv,L.Chen2022JSAC,M.Temiz2021TCCN,X.Wang2022CL}. (a) Integration of sensing with downlink communication: An ISAC signal is sent by the BS to perform simultaneous downlink communication and radar sensing, and the receive side of the BS remains active for the reception of radar echo;
(b) Integration of sensing with uplink communication: The BS transmits a pure sensing signal and receives the echoes during the uplink communication.}\label{fig:ISAC_existing}
\end{figure}

\begin{table*}[t]
\caption{Comparisons With Previous FD ISAC Works}
    \centering
    \setlength\tabcolsep{1.5mm}{
	\begin{tabular}{|c|c|c|c|c|}
\hline  \multirow{2}*{\textbf{Study}} & \multicolumn{3}{|c|}{\textbf{System model}}& \multirow{2}*{\textbf{Main objective}} \\
\cline{2-4}  & FD rad & DL com & UL com &   \\
\hline  \cite{Liu2022CRB,C.Tsinos2021JSTSP,J.Pritzker2022Arxiv,L.Chen2022JSAC} & $\ast$ & $\ast$ &  &  \makecell[l]{Optimize the transmit signal and radar receive beamformer of downlink ISAC for different metrics} \\
\hline  \cite{M.Temiz2021TCCN,X.Wang2022CL}& $\ast$ &  & $\ast$ & \makecell[l]{Investigate the receive strategy with concurrent communication and sensing signal reception}\\
\hline  \cite{FDISAC}& $\ast$ &$\ast$ & $ $ & \makecell[l]{Study and measure the sensing performance of LTE and 5G NR waveforms and provide the SIC strategies}\\
\hline  This work& $\ast$ &$\ast$ & $\ast$ & \makecell[l]{Deal with the joint optimization for general ISAC involved FD communication and sensing}\\
\hline
\multicolumn{5}{l}{{\textit{Notes:} DL, UL, rad, and com represent downlink, uplink, radar, and communication, respectively.}}
	\end{tabular}%
}
\end{table*}\label{table1}

The main function of a radar system is to estimate the channel parameters, e.g., delay and Doppler frequency, of a target from the received radar echo signal.
With the consideration of radar echo reception in ISAC systems, e.g., \cite{Liu2022CRB,C.Tsinos2021JSTSP,J.Pritzker2022Arxiv,L.Chen2022JSAC,M.Temiz2021TCCN,X.Wang2022CL}, the associated scenarios are divided into two cases, as illustrated in Fig. \ref{fig:ISAC_existing}. The first case in Fig. \ref{fig:ISAC_existing}(a) corresponds to downlink ISAC \cite{Liu2022CRB,C.Tsinos2021JSTSP,J.Pritzker2022Arxiv,L.Chen2022JSAC}, where the radar sensing reuses the resources of downlink transmission and the BS acts as a radar transceiver and a communication transmitter. The transmitted downlink ISAC signal is known to the BS and can be used in receive processing for sensing.
In this case, the authors of \cite{Liu2022CRB} investigated the Cram\'er-Rao bound (CRB) minimization of target parameter estimation for ISAC. In \cite{C.Tsinos2021JSTSP,J.Pritzker2022Arxiv,L.Chen2022JSAC}, the authors considered the tasks of point target detection in ISAC systems. In these works, they acquired explicitly the radar SINR for target detection by applying a linear receive beamformer to the echo signal.
More concretely, given a fixed radar receive beamformer, the optimizations of the transmit signals were investigated in \cite{J.Pritzker2022Arxiv} and \cite{L.Chen2022JSAC}, where a minimal radar SINR requirement for accomplishing the target detection is constrained.
In \cite{C.Tsinos2021JSTSP}, an alternating optimization (AO)-based algorithm was proposed to iteratively update the transmit waveform and the radar receive beamformer.
The second scenario in Fig. \ref{fig:ISAC_existing}(b) considers integrating sensing with uplink communication \cite{M.Temiz2021TCCN,X.Wang2022CL}, where the BS can be regarded as a radar transceiver and a communication receiver.
The authors of \cite{M.Temiz2021TCCN} developed an advanced receiver architecture for uplink ISAC, which separates the radar echo and communication signals by performing interference cancellation techniques.
Sensing-assisted physical-layer security transmission was investigated in \cite{X.Wang2022CL}, where the BS transmits a downlink radar signal to localize and jam a potential aerial eavesdropper while receiving the uplink communication signal.

In the aforementioned works \cite{Liu2022CRB,C.Tsinos2021JSTSP,J.Pritzker2022Arxiv,L.Chen2022JSAC,M.Temiz2021TCCN,X.Wang2022CL}, the radar receiver operates simultaneously while transmitting, i.e., in a full-duplex (FD) manner \cite{J.Zhang2022CSTEnabling,C.B.BarnetoWCM2021Full,FDISAC}.
In particular, self-interference (SI), which is a critical issue in FD operation, is considered to be suppressed by employing advanced SI cancellation (SIC) techniques \cite{FDSI1,FDSI5}, e.g., natural isolation, analog cancellation, and digital cancellation.
Particularly, in an FD ISAC system, the SIC should be performed only for the direct signal coupling between the transceiver antennas, while the target reflections should be preserved \cite{C.B.BarnetoWCM2021Full,FDISAC}.
With FD radar, however, the integrated communication functionality occurs only in either the downlink or the uplink, operating in a half-duplex (HD) manner \cite{Liu2022CRB,C.Tsinos2021JSTSP,J.Pritzker2022Arxiv,L.Chen2022JSAC,M.Temiz2021TCCN,X.Wang2022CL}.
Therefore, to achieve higher spectral efficiency, it is motivated to consider the FD capability also for communication \cite{FDSI1,FDSI5}, i.e., to let the BS serve as both a radar transceiver and a communication transceiver concurrently.
Under this setup, there is not only interference between sensing and communication functionalities, but also coupling between uplink and downlink transmissions, that significantly complicate the ISAC design.
Existing algorithms in \cite{Liu2022CRB,C.Tsinos2021JSTSP,J.Pritzker2022Arxiv,L.Chen2022JSAC,M.Temiz2021TCCN,X.Wang2022CL} cannot be straightforwardly applied to address these challenges.
Specifically, the algorithms designed in \cite{Liu2022CRB,C.Tsinos2021JSTSP,J.Pritzker2022Arxiv,L.Chen2022JSAC} do not incorporate the impact of uplink communication. In \cite{M.Temiz2021TCCN,X.Wang2022CL}, only a pure downlink sensing signal is sent and the uplink transmit power is fixed, without considering the possibility of downlink communication nor designing the uplink transmission.

Motivated by the above discussions, we investigate an advanced FD communication-based ISAC system, where the BS receives and transmits signals from multiple uplink users and downlink users reusing the same time and frequency resources. The downlink transmit signal is an ISAC signal that is applied for both conveying information to the downlink users and performing a sensing task of point target detection. The BS also simultaneously conducts uplink communication signal reception and processes the radar echo signal. Our goal is to jointly design the transceiver beamforming at the FD BS and the transmit power at the single-antenna uplink users. In the considered FD system, the SI at the BS, or more precisely, the direct signal coupling link between the transceiver, is assumed to be suppressed to an acceptable level by employing SIC techniques for ISAC systems \cite{C.B.BarnetoWCM2021Full,FDISAC,FDSI1,FDSI5}.
A brief comparison of this study with previous FD ISAC works is summarized in Table~\ref{table1}.

In the considered FD ISAC systems, the received signals at the BS consist of uplink communication signals, desired target reflection, and downlink signal-dependent interference from environmental interferers and residual SI.
To detect both the sensing target and multiuser uplink signals with low complexity, multiple linear receive beamformers are employed at the BS and the corresponding radar and uplink communication SINRs are mathematically obtained.
As such, we formulate two different fundamental problems for the joint optimization.
The first problem focuses on power minimization by constraining the minimal SINR requirements of target detection, uplink communications, and downlink communications. The second problem aims at maximizing the sum rate of the FD multiuser communication, subject to the constraint of minimal sensing SINR requirement and the limit of maximal transmit powers.
Compared to \cite{Liu2022CRB,C.Tsinos2021JSTSP,J.Pritzker2022Arxiv,L.Chen2022JSAC,M.Temiz2021TCCN,X.Wang2022CL}, we consider the joint optimization for both uplink and downlink transmissions of the FD ISAC system, which are highly coupled and intractable.
The main contributions of this paper are summarized as follows:
\begin{itemize}
\item We extend the existing ISAC beamforming design, e.g., \cite{Liu2022CRB,C.Tsinos2021JSTSP,J.Pritzker2022Arxiv,L.Chen2022JSAC,M.Temiz2021TCCN,X.Wang2022CL}, to a general case by considering the FD capability for not only sensing but also communication and focusing on the optimization of coupled downlink and uplink transmissions. With the employment of linear receive beamformers, the SINRs of radar sensing and communication of the FD ISAC system are mathematically formulated and two different problems are constructed aiming to improve the system power efficiency and spectral efficiency, respectively.
  \item We derive the optimal receive beamformers to maximize the SINR of target detection and the SINRs of uplink communication, respectively, which are obtained as closed-form expressions with respect to the BS transmit beamforming and the user transmit power.
  \item For each of the two considered problems, we first obtain an equivalent problem that involves the optimization of only the BS transmit beamforming and the user transmit power based on the closed-form receivers. Subsequently, an iterative algorithm is proposed to find a high-quality solution by applying the techniques of rank relaxation and successive convex approximation (SCA).
      Moreover, we prove that the adopted relaxation is tight.
  \item For the problem of power minimization, we further consider a special case of HD uplink communication-based ISAC in the absence of downlink users, while the downlink signal is adopted for target detection only.
      Instead of applying the SCA-based algorithm as in the general case, we propose an AO-based algorithm to iteratively update the receive beamformers and the other variables,
      whose solutions are obtained by calculating closed-form expressions and by solving a second-order cone programming (SOCP), respectively. Numerical results verify that this newly proposed method significantly reduces the computational complexity compared to the SCA-based method with almost the same performance.
\end{itemize}

The rest of this paper is organized as follows. In Section II, we present the model of the considered FD ISAC system and the formulations of the power minimization and sum rate maximization problems. Section III and Section IV provide detailed solutions to these two problems, respectively. In Section V, we verify the effectiveness of the proposed algorithms through numerical simulations. Conclusions are drawn in Section~VI.

\textit{Notations:}
Boldface lower-case and boldface upper-case letters are used to represent vectors and matrices, respectively. Denote superscripts $(\cdot)^T$ and $(\cdot)^H$ by the transpose and the Hermitian transpose, respectively. Let $\text{Tr}(\cdot)$, $\text{rank}(\cdot)$, and $[\cdot]_{i,j}$ return the trace, the rank, and the $(i,j)$-th entry of a matrix, respectively. Denote $\|\cdot\|$ by the $\ell_2$ norm of a vector and $|\cdot|$ by the absolute value of a scalar.
We use $\mathbb E\{ \cdot \}$ for the expectation operation, $\mathcal I\{\cdot \}$ for the imaginary part of a complex-valued number, $\mathbb C$ for the set of complex-value numbers, and $\mathbf I_{N}$ for the identity matrix of size $N \times N$.
Let $\mathbf X \succeq \mathbf 0$ imply that $\mathbf X$ is positive semidefinite. Denote $\mathcal O(\cdot)$ by the big-O computational complexity notation.

\begin{figure}[t]
 \begin{center}
      \epsfxsize=7.0in\includegraphics[scale=0.6]{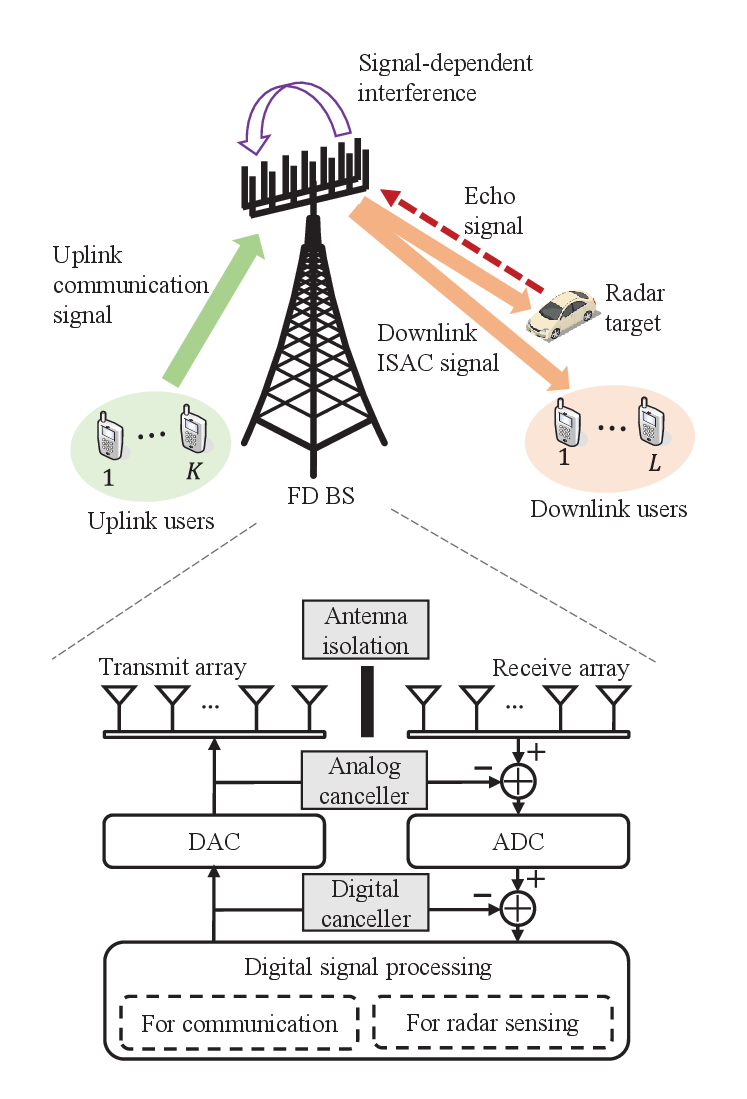}
      \caption{The considered FD communication-based ISAC system with $K$ uplink users, $L$ downlink users, and a point radar target.}\label{fig:sysmodel}
    \end{center}
\end{figure}

\section{System Model and Problem Formulation}
Consider an ISAC system as shown in Fig. \ref{fig:sysmodel}, where a dual-functional FD BS equipped with two uniform linear arrays (ULAs) receives the communication signals from $K$ single-antenna uplink users and sends a downlink ISAC signal via the same time-frequency resource.
A diagram of the architectural procedure for the SIC that can be used in an FD ISAC system is also included in Fig. \ref{fig:sysmodel}.
The downlink ISAC signal transmitted from an $N_t$-element ULA is adopted for simultaneously communicating with $L$ single-antenna downlink users and performing target detection on a point radar target.
The radar echo signal and the uplink communication signals are received at the BS through the receive ULA with $N_r$ elements.

The FD operation is beneficial to both communication and sensing functionalities by reusing the time-frequency resource efficiently. From the communication perspective, the spectral efficiency is significantly improved. Meanwhile from the radar perspective, the sensing is continuously performed at the BS occupying all the available channel bands such that an enhanced radar performance is achieved \cite{J.Zhang2022CSTEnabling,C.B.BarnetoWCM2021Full,RadarBook}.

\subsection{Signal Model}
We first focus on the downlink transmission of the system, where a narrowband ISAC signal, $\mathbf x \in \mathbb C^{N_t \times 1}$, is sent for simultaneous radar sensing and downlink multiuser communication via multi-antenna beamforming.
Following \cite{XLiuTSP2020,HHuaArxiv,Z.LyuArxiv,Z.HeWCL2022,Liu2022CRB}, the integrated signal is expressed as
\begin{align}\label{signal:xISAC}
\mathbf x = \sum_{l=1}^L \mathbf v_l s_l + \mathbf s_0,
\end{align}
where $\mathbf v_l \in \mathbb C^{N_t \times 1}$ stands for the beamforming vector associated with downlink user $l$, $l \in \{1,\cdots,L\},$ and $s_l \in \mathbb C$ is the data symbol of user $l$ with unit power, i.e., $\mathbb{E}\{ |s_l|^2 \} = 1$. Here, $\mathbf s_0 \in \mathbb C^{N_t \times 1}$ represents a dedicated radar signal with covariance matrix $\mathbf V_0 \triangleq \mathbb{E}\{\mathbf s_0 \mathbf s_0^H\}$, for extending the DoF of the transmit signal $\mathbf x$ to achieve enhanced sensing performance~\cite{XLiuTSP2020}. The signals $\{s_l\}_{l=1}^L$ and $\mathbf s_0$ are assumed to be independent with each other. In (\ref{signal:xISAC}), the downlink beamforming is achieved by designing $\{ \mathbf v_l\}_{l=1}^L$ and $\mathbf V_0$ \cite{XLiuTSP2020,HHuaArxiv,Z.LyuArxiv,Z.HeWCL2022,Liu2022CRB}.
Once $\mathbf V_0$ is determined, the dedicated radar signal $\mathbf s_0$ can be generated \cite{P.StoicaTSP2007}.
Moreover, we consider a total transmit power constraint as
$\sum_{l=1}^L \| \mathbf v_l \|^2 + \text{Tr}(\mathbf V_0) \leq P_\text{max}$, where $P_\text{max}$ denotes the maximum available power budget of the BS.

When the FD BS transmits $\mathbf x$, it simultaneously receives the uplink communication signal and the target reflection. Let $d_k \in \mathbb C$ denote the uplink signal from user $k$, $k \in \{1,\cdots,K\}$, which satisfies
\begin{align}
\mathbb E\{|d_k|^2\} = p_k,\ \forall k,
\end{align}
where $0 \leq  p_k\leq P_k $ represents the average transmit power of user $k$ with $P_k$ being the maximum power budget.
Denoting the uplink channel between the $k$-th user and the BS by $\mathbf h_k \in \mathbb C^{N_r \times 1}$,
the received multiuser uplink signal at the BS is $\sum_{k=1}^K \mathbf h_k d_k$. The design of uplink transmission is achieved by adjusting the transmit power $\{p_k\}_{k=1}^K$ of uplink users.

We next model the echo signal of the considered MIMO radar.
Assume that the radar channel consists of line-of-sight (LoS) paths and both the transmit and receive ULAs at the BS are half-wavelength antenna spacing. We denote the transmit array steering vector to direction $\theta$ by
$\mathbf a_t(\theta) \triangleq \frac{1}{\sqrt{N_t}}  [1,e^{j\pi\sin(\theta)}, \cdots, e^{j\pi(N_t-1)\sin(\theta)}]^T$ and similarly denote by $\mathbf a_r(\theta) \triangleq \frac{1}{\sqrt{N_r}} [1,e^{j\pi\sin(\theta)}, \cdots, e^{j\pi(N_r-1)\sin(\theta)}]^T$ the receive steering vector.
Supposing that the target to be detected is located at angle $\theta_0$, the target reflection is given by $\beta_0 \mathbf a_r(\theta_0) \mathbf a_t^H(\theta_0) \mathbf x$, where $\beta_0 \in \mathbb C $ is the complex amplitude of the target mainly determined by the path loss and the radar cross-section \cite{J.Pritzker2022Arxiv}.
We assume that $\theta_0$ and $\beta_0$ are known or previously estimated at the BS for designing the best suitable transmit signal to detect this specific target of interest, like in \cite{L.Chen2022JSAC,C.Tsinos2021JSTSP,C.Y.ChenTSP2009MIMOradar,Wu2018Transmit,radarCM}.
Based on the given uplink communication signal and the target echo, we express the received signal at the FD BS as
\begin{align}\label{echo}
\mathbf y^\text{BS} =& \sum_{k=1}^K \mathbf h_k d_k + \beta_0 \mathbf A(\theta_0) \mathbf x + \mathbf z + \mathbf n,
\end{align}
where $\mathbf A(\theta_0) \triangleq \mathbf a_r (\theta_0) \mathbf a_t^H(\theta_0)$, $\mathbf n \in \mathbb C^{N_r \times 1}$ denotes additive white Gaussian noise (AWGN) with covariance $\sigma_r^2 \mathbf I_{N_r}$, and $\mathbf z \in \mathbb C^{N_r \times 1}$ represents the undesired signal-dependent interference which is detailed in the next paragraph.

The signal-dependent interference $\mathbf z$ can be decomposed into two parts. The first part corresponds to the clutter reflected from the surrounding environment. Without loss of generality, we follow \cite{C.Tsinos2021JSTSP,Wu2018Transmit} and assume that there exist $I$ signal-dependent uncorrelated interferers located at angles $\{ \theta_i\}_{i=1}^I$ and $\theta_i \neq \theta_0,\ \forall i \in \{1,\cdots,I\}$. These $I$ interferers also reflect the sensing signal to the BS, yielding the undesired interference $\sum_{i=1}^I \beta_i \mathbf A(\theta_i) \mathbf x$ with $\beta_i \in \mathbb C $ being the complex amplitude of the $i$-th interferer and $\mathbf A(\theta_i) \triangleq \mathbf a_r (\theta_i) \mathbf a_t^H(\theta_i),\ \forall i$. The second part is the SI caused by the considered FD operation. By employing SIC techniques for ISAC systems \cite{C.B.BarnetoWCM2021Full,FDISAC}, the SI power can be mostly reduced.
Without loss of generality, we take a general model to express the residual SI signal as $\mathbf H_\text{SI} \mathbf x$ \cite{FDSI1}, where $\mathbf H_\text{SI} \in \mathbb C^{N_r \times N_t}$ denotes the residual SI channel at the FD BS.
Combining the two parts of interference, $\mathbf z$ is expressed as
\begin{align}\label{def:d}
\mathbf z = \sum_{i=1}^I \beta_i \mathbf A(\theta_i) \mathbf x + \mathbf H_\text{SI} \mathbf x.
\end{align}

By substituting (\ref{def:d}) into (\ref{echo}), we express the complete received signal at the FD BS as
\begin{align}\label{signal:y}
\mathbf y^\text{BS} = \!\!\!\!\!\! \underbrace{\sum_{k=1}^K \mathbf h_k d_k}_\text{Communication signal}\!\!\!\!\!\! + \underbrace{ \beta_0 \mathbf A(\theta_0) \mathbf x}_\text{Target reflection} + \!\!\! \!\!\!\underbrace{\sum_{i=1}^I \beta_i \mathbf A(\theta_i) \mathbf x}_\text{Echo signal of interferers} \!\!\!\!\!\! +  \underbrace{\mathbf H_\text{SI} \mathbf x}_\text{SI} + \mathbf n.
\end{align}
On the other hand, denote the channel between downlink user $l$ and the BS by $\mathbf g_l \in \mathbb C^{N_t\times 1}$. The received signal at downlink user $l$ is then expressed as
\begin{align}\label{signal:yuser}
y^\text{User}_l = \!\!\! \underbrace{\mathbf g_l^H \mathbf v_l s_l }_\text{Desired signal} \! + \!\!\underbrace{\sum_{l'=1,l'\neq l}^L \mathbf g_l^H \mathbf v_{l'} s_{l'}}_\text{Multiuser interference} + \!\!\! \underbrace{\mathbf g_l^H \mathbf s_0}_\text{Sensing signal}\!\!\! +\ n_l,\ \forall l,
\end{align}
where $n_l$ stands for the AWGN with variance $\sigma^2_l$.
Although a few prior works on the uplink ISAC, e.g., \cite{X.Wang2022CL,M.Temiz2021TCCN}, also considered a similar received signal model at the ISAC BS as (\ref{signal:y}), this paper is noticeably different from these works in terms of the following two aspects. First, the downlink transmitted $\mathbf x$ in our system is a dual-functional ISAC signal that serves both communication and sensing, while only a pure radar signal was sent in \cite{X.Wang2022CL,M.Temiz2021TCCN}. Second, studies \cite{X.Wang2022CL,M.Temiz2021TCCN} mainly focused on the receiver design with a fixed uplink transmit power without adaptability. To enable more design flexibility and gain further performance improvements, in this paper, we additionally introduce the optimization of the uplink transmission, i.e., $\{p_k\}_{k=1}^K$.

Before proceeding, we would like to clarify some assumptions employed in the considered system. First in (\ref{echo}), the angles of a target seen at the transceiver are identical, which is a reasonable assumption when the transmit array and the receive array are colocated \cite{P.StoicaMIMOradar1}. Second, it is assumed that $\{\theta_i\}_{i=1}^I$ and $\{\beta_i\}_{i=1}^I$ in (\ref{def:d}) can be pre-estimated and known to the ISAC system for transceiver design \cite{C.Tsinos2021JSTSP} by using an environmental dynamic database \cite{Wu2018Transmit}. Finally, we assume that a dedicated channel estimation stage is utilized before the FD transmission such that the channel state information (CSI) is available at the BS for beamforming design\cite{M.Temiz2021TCCN}.

\subsection{Radar and Communication SINR}
The performances of the radar and the communication systems largely depend on the corresponding SINRs.
In particular, when considering point target detection in MIMO radar systems, the detection probability of a target is generally a monotonically increasing function of the output SINR \cite{radarCM}. Therefore, we directly adopt radar SINR as the performance metric of the sensing functionality.
Technically, we apply a receive beamformer $\mathbf u \in \mathbb C^{N_r \times 1}$ on the received signal, $\mathbf y^\text{BS}$, to capture the desired reflected signal of the point target. Then, based on (\ref{signal:y}), we obtain the radar SINR as
\begin{align}\label{sinr_r}
\gamma^\text{rad}
=& \frac{\mathbb E \{ | \mathbf u^H \beta_0 \mathbf A(\theta_0) \mathbf x |^2 \}}{\sum_{k=1}^K \mathbb E \{ |  \mathbf u^H \mathbf h_{k} d_{k}|^2 \}  +  \mathbb E \{| \mathbf u^H \mathbf B \mathbf x |^2\} + \mathbb E \{| \mathbf u^H \mathbf n |^2\}} \nonumber \\
=& \frac{| \beta_0|^2 \mathbf u^H \mathbf A(\theta_0) \mathbf Q \mathbf A(\theta_0)^H \mathbf u}
{\mathbf u^H \left( \sum_{k=1}^K p_k  \mathbf h_k \mathbf h_k^H +  \mathbf B \mathbf Q {\mathbf B}^H + \sigma^2_r \mathbf I_{N_r} \right)\mathbf u},
\end{align}
where $\mathbf B \triangleq \sum_{i=1}^I \beta_i \mathbf A(\theta_i) + \mathbf H_\text{SI}$ represents the interference channel defined as the summation of the $I$ interferers' channels and the SI channel, and
\begin{align}
\mathbf Q \triangleq \mathbb E \{ \mathbf x \mathbf x^H\} =\sum_{l=1}^L \mathbf v_l \mathbf v_l^H + \mathbf V_0
\end{align}
denotes the covariance matrix of the downlink ISAC signal that needs to be well designed.
Note that the radar SINR was employed as a sensing metric for beamforming design in ISAC before, but only under the downlink scenario in the absence of uplink communication, e.g., \cite{J.Pritzker2022Arxiv,L.Chen2022JSAC,C.Tsinos2021JSTSP}.
The term $\sum_{k=1}^K p_k  \mathbf h_k \mathbf h_k^H + \mathbf B \mathbf Q {\mathbf B}^H$ makes the radar SINR expression in (\ref{sinr_r}) in our considered FD ISAC more complicated than those in \cite{J.Pritzker2022Arxiv,L.Chen2022JSAC,C.Tsinos2021JSTSP}, as it introduces signal-dependent interference and coupled uplink transmission.

Similarly, by applying another set of receive beamformers $\{\mathbf w_k\}_{k=1}^K \in \mathbb C^{N_r \times 1}$ on $\mathbf y^\text{BS}$ to recover the data signals of the uplink users, we obtain the corresponding receive SINR of user $k$ by
\begin{align}\label{sinr_k}
\gamma_k^\text{com,UL}\!\!
=\!\! \frac{p_k \mathbf w_k^H \mathbf h_k \mathbf h_k^H \mathbf w_k}
{\mathbf w_k^H \!\! \left( \! \sum_{k'=1, k' \neq k}^K \! p_{k'}  \mathbf h_{k'} \mathbf h_{k'}^H \!\!\! +\!  \mathbf C \mathbf Q \mathbf C^H \!\! +\! \sigma_r^2 \mathbf I_{N_r} \! \right)\! \mathbf w_k},\ \forall k,
\end{align}
where $\mathbf C \triangleq \sum_{i=0}^I \beta_i \mathbf A(\theta_i) + \mathbf H_\text{SI}$ denotes the interference channel caused by the downlink transmission.
As for the downlink communication, it follows from (\ref{signal:yuser}) that the SINR of the downlink user $l$ is given by
\begin{align}\label{sinr_l}
\gamma_l^\text{com,DL} =& \frac{|\mathbf g_l^H \mathbf v_l |^2}{\sum_{l'=1,l'\neq l}^L |\mathbf g_l^H \mathbf v_{l'} |^2 + \mathbf g_l^H \mathbf V_0 \mathbf g_l  + \sigma^2_l},\ \forall l.
\end{align}
Here, we consider the users which are not capable of canceling the interference from the dedicated radar signal $\mathbf s_0$. Usually when interference cancellation schemes are employed at users \cite{HHuaArxiv}, we only need to remove the interference term $\mathbf g_l^H \mathbf V_0 \mathbf g_l$ from the denominator in (\ref{sinr_l}) and the proposed algorithms remain applicable.
In addition, in this paper we mainly focus on a general transceiver beamforming design, without imposing any strict constraints or strategies of nulling the interferences involved in the system, e.g., the sensing-communication interference in (\ref{sinr_r}) and (\ref{sinr_k}) and the multiuser interference in (\ref{sinr_k}) and (\ref{sinr_l}).
However, through the beamforming optimization, these interferences can be somewhat suppressed and a superior system performance can be achieved.

\subsection{Problem Formulation}
We aim at jointly optimizing the transmit power, $\{p_k\}_{k=1}^K$, at the uplink users, the receive beamformers, $\{ \mathbf w_k\}_{k=1}^K$ and $\mathbf u$, and the transmit beamforming, $\{\mathbf v_l\}_{l=1}^L$ and $\mathbf V_0$, at the BS for the considered FD ISAC system. Denote $\mathcal A \triangleq \left\{\{ \mathbf w_k\}_{k=1}^K,\mathbf u,\{\mathbf v_l\}_{l=1}^L,\mathbf V_0\succeq \mathbf 0, \{p_k\geq 0\}_{k=1}^K \right\}$ as the set of optimization variables.
The joint design is performed under two criteria: 1) transmit power minimization; 2) overall sum rate maximization, which correspond to the power efficiency and the spectral efficiency improvement of the ISAC system, respectively.
Specifically, for the first design criterion, we consider minimizing the total transmit power consumption while guaranteeing the minimal SINR requirements of uplink communications, downlink communications, and radar sensing.
The corresponding problem is formulated as
\begin{align}\label{prob:minP}
\mathop \text{minimize} \limits_{\mathcal A} \quad
& \sum_{l=1}^L \| \mathbf v_l \|^2 + \text{Tr}(\mathbf V_0) + \sum_{k=1}^K p_k \nonumber \\
\text{subject to}\quad
& \gamma^\text{rad} \geq \tau^\text{rad}, \nonumber \\
& \gamma_k^\text{com,UL} \geq \tau_k^\text{com,UL}, \ \forall k, \nonumber \\
& \gamma_l^\text{com,DL} \geq \tau_l^\text{com,DL}, \ \forall l,
\end{align}
where $\tau^\text{rad}$ is the required constant minimal SINR threshold for successfully accomplishing the sensing operation, and $\tau_k^\text{com,UL}$ and $\tau_l^\text{com,DL}$ stand for the minimal SINR requirements of uplink user $k$ and downlink user $l$, respectively.

We also wish to maximize the sum rate of all the uplink and downlink users with limited transmit power budgets, while ensuring the sensing performance by constraining the minimal radar SINR. Accordingly, we formulate the problem as
\begin{align}\label{prob:maxR}
\mathop \text{maximize} \limits_{\mathcal A} \quad
& \sum_{k=1}^K \log_2(1 + \gamma_k^\text{com,UL}) + \sum_{l=1}^L \log_2(1 + \gamma_l^\text{com,DL}) \nonumber \\
\text{subject to}\quad
& \gamma^\text{rad} \geq \tau^\text{rad}, \nonumber \\
& \sum_{l=1}^L \| \mathbf v_l \|^2 + \text{Tr}(\mathbf V_0) \leq P_\text{max}, \nonumber \\
& p_k \leq P_k,\ \forall k.
\end{align}

Observe that both (\ref{prob:minP}) and (\ref{prob:maxR}) are nonconvex problems whose globally optimal solutions are hard to obtain by polynomial-time algorithms in general. Moreover, the optimization variables are tightly coupled which further complicates the problems and makes them intractable.
In the following sections, we propose efficient algorithms to solve these two problems, respectively.

\begin{remark}
Note that the sensing task can be extended to multi-target scenarios with $M\geq 2$ targets. Specifically, by utilizing $M$ radar receive beamformers $\{\mathbf u_m\}_{m=1}^M$ to process the received signal at the BS, the SINRs $\{ \gamma^\text{rad}_m\}_{m=1}^M$ for all the $M$ targets can be separately obtained \cite{Multitarget}. The single radar SINR constraint involved in problems (\ref{prob:minP}) and (\ref{prob:maxR}) is then replaced with $M$ individual constraints as $\gamma^\text{rad}_m \geq \tau^\text{rad}_m,\ \forall m$.
\end{remark}

\section{Joint FD ISAC Design for Power Minimization}
We handle (\ref{prob:minP}) in this section.
Specifically, we first determine the optimal receive beamformers $\left\{\{\mathbf w_k\}_{k=1}^K, \mathbf u\right\}$ in closed-form expressions with respect to $\left\{\{\mathbf v_l\}_{l=1}^L, \mathbf V_0, \{p_k \}_{k=1}^K\right\}$ and substitute them into (\ref{prob:minP}). Then, we address the equivalent problem exploiting the SCA technique. In addition, we investigate a special case and provide a low-complexity solution.

\subsection{Closed-Form Solutions to Receive Beamformer}
Note that the objective of (\ref{prob:minP}) does not depend on $\left\{\{\mathbf w_k\}_{k=1}^K, \mathbf u\right\}$. Moreover, given arbitrary feasible $\left\{\{\mathbf v_l\}_{l=1}^L,\mathbf V_0,\{p_k \}_{k=1}^K \right\}$, it can be found that $\mathbf u$ only affects the value of $\gamma^\text{rad}$ and $\mathbf w_k$ has an impact on $\gamma^\text{com,UL}_k$ while it does not affect the SINRs of other users.
Therefore, to facilitate the fulfillment of the SINR constraints in (\ref{prob:minP}) and reduce the transmit power consumption, the receive beamformers $\left\{\{\mathbf w_k\}_{k=1}^K, \mathbf u\right\}$ should be determined by maximizing the corresponding SINRs.
As a result, we optimize them through the SINR maximization criterion:
\begin{align}
\mathop \text{maximize}\limits_{ \mathbf u} \quad & \gamma^\text{rad}, \label{prob:w_r}\\
\mathop \text{maximize}\limits_{ \mathbf w_k} \quad & \gamma^\text{com,UL}_k , \ \forall k . \label{prob:w_k}
\end{align}

\begin{proposition}\label{prop:receiver}
The optimal solutions to (\ref{prob:w_r}) and (\ref{prob:w_k}) are given by
\begin{align}
\mathbf u^* =& \left( \sum_{k=1}^K p_k \mathbf h_k \mathbf h_k^H + \mathbf B \mathbf Q \mathbf B^H + \sigma_r^2 \mathbf I_{N_r} \right)^{-1}\!\!\! \mathbf a_r (\theta_0), \label{w_r*}\\
\mathbf w_k^* =& \left(\! \sum_{k'=1, k'\neq k}^K \!\!\! p_{k'}  \mathbf h_{k'} \mathbf h_{k'}^H + \mathbf C \mathbf Q \mathbf C^H + \sigma_r^2 \mathbf I_{N_r} \right)^{-1}\!\!\! \mathbf h_k, \ \forall k, \label{w_k*}
\end{align}
respectively, which rely on the transmit beamforming at the BS and the transmit power of uplink users.
\end{proposition}
\begin{IEEEproof}
See Appendix \ref{proof:propreceiver}.
\end{IEEEproof}
Note that scaling $\mathbf u^*$ and $\mathbf w_k^*,\ \forall k,$ with any positive constant does not affect the optimality.

\subsection{Solutions to Transmit Beamforming and Power}
Substituting the optimal $\left\{\{\mathbf w_k^*\}_{k=1}^K, \mathbf u^*\right\}$ into the SINR expressions in (\ref{sinr_r}) and (\ref{sinr_k}) yields
\begin{align}
&\bar{\gamma}^\text{rad} \!\!= |\beta_0|^2 \mathbf a_t^H(\theta_0) \mathbf Q \mathbf a_t (\theta_0)  \nonumber\\
&\ \quad\quad \times \!\mathbf a_r^H\!(\theta_0)\!\! \left(\sum_{k=1}^K p_k  \mathbf h_k \mathbf h_k^H \!\!\!+\!  \mathbf B \mathbf Q \mathbf B^H \!\!\!+\! \sigma_r^2 \mathbf I_{N_r} \!\!\right)^{-1}\!\!\!\!\!\! \mathbf a_r\!(\theta_0), \label{sinr_r'}\\
&\bar{\gamma}_k^\text{com,UL}\!\! = p_k \mathbf h_k^H \!\! \left(\!\sum_{k'=1, k'\neq k}^K \!\!\!\!\!p_{k'}  \mathbf h_{k'} \mathbf h_{k'}^H \!\!+\!  \mathbf C \mathbf Q \mathbf C^H \!\!+\! \sigma_r^2 \mathbf I_{N_r} \!\!\right)^{-1}\!\!\!\!\!\!\! \mathbf h_k, \ \forall k, \label{sinr_k'}
\end{align}
respectively. Subsequently, applying (\ref{sinr_r'}) and (\ref{sinr_k'}), we rewrite (\ref{prob:minP}) into the following equivalent problem with respect to $\left\{\{\mathbf v_l\}_{l=1}^L, \mathbf V_0, \{p_k \}_{k=1}^K\right\}$:
\begin{align}\label{prob:minPtilde}
\mathop \text{minimize} \limits_{\{\mathbf v_l\}_{l=1}^L,\mathbf V_0 \succeq \mathbf 0, \atop \{p_k \geq 0 \}_{k=1}^K } \
& \sum_{l=1}^L \| \mathbf v_l \|^2 + \text{Tr}(\mathbf V_0) + \sum_{k=1}^K p_k \nonumber \\
\text{subject to}\quad
& \bar{\gamma}^\text{rad} \geq \tau^\text{rad}, \nonumber \\
& \bar{\gamma}_k^\text{com,UL} \geq \tau_k^\text{com,UL}, \ \forall k, \nonumber \\
& \gamma_l^\text{com,DL} \geq \tau_l^\text{com,DL}, \ \forall l.
\end{align}
The above problem is still nonconvex due to the complicated SINR constraints. To handle this issue,
we introduce a set of auxiliary variables $\mathbf V_l \triangleq \mathbf v_l \mathbf v_l^H,\ \forall l$. With $\{\mathbf V_l\}_{l=1}^L$, we further define $\mathbf {\bar Q} \triangleq \sum_{l=0}^L \mathbf V_l$, $\mathbf \Psi \triangleq \sum_{k=1}^K p_k  \mathbf h_k \mathbf h_k^H +  \mathbf B \mathbf {\bar Q} \mathbf B^H + \sigma_r^2 \mathbf I_{N_r}$, and $\mathbf \Phi_k
\triangleq \sum_{k'=1,k'\neq k}^K p_{k'}  \mathbf h_{k'} \mathbf h_{k'}^H +\mathbf C \mathbf {\bar Q} \mathbf C^H + \sigma_r^2 \mathbf I_{N_r}, \ \forall k,$ to simplify the SINR expressions.
After some straightforward algebraic operations, (\ref{prob:minPtilde}) is recast as
\begin{align}
\mathop \text{minimize} \limits_{\{\mathbf V_l \succeq \mathbf 0 \}_{l=0}^L, \atop \{p_k \geq 0 \}_{k=1}^K } \quad
& \sum_{l=0}^L \text{Tr}(\mathbf V_l)+ \sum_{k=1}^K p_k \label{prob:minPtildeRef}  \\
\text{subject to}\quad
& \mathbf a_t^H\!(\theta_0\!) \mathbf {\bar Q} \mathbf a_t(\theta_0) \mathbf a_r^H\!(\theta_0\!)\mathbf \Psi^{-1} \mathbf a_r(\theta_0 \!) \!\geq\! \frac{\tau^\text{rad}}{|\beta_0|^2},                                 \tag{\ref{prob:minPtildeRef}{a}} \label{cons:minPrad} \\
& \mathbf h_k^H  \mathbf \Phi_k^{-1} \mathbf h_k \geq \frac{\tau_k^\text{com,UL}}{p_k}, \ \forall k, \tag{\ref{prob:minPtildeRef}{b}} \label{cons:minPUL} \\
& \left(1+\frac{1}{\tau_l^\text{com,DL}} \right)\mathbf g_l^H \mathbf V_l \mathbf g_l \geq \mathbf g_l^H \mathbf {\bar Q}\mathbf g_l+ \sigma^2_l,  \nonumber\\
&\quad\quad \quad \quad \quad \quad \quad \quad \quad \quad \quad \quad \quad \forall l \geq 1.  \tag{\ref{prob:minPtildeRef}{c}} \label{cons:minPDL}
\end{align}
Note that we omitted the rank constraints of $\{\mathbf V_l\}_{l=1}^L$, i.e.,
\begin{align}\label{cons:rank1}
\text{rank}(\mathbf V_l) \leq 1, \ \forall l\geq 1,
\end{align}
based on the idea of rank relaxation \cite{GaussianR}.
The reformulation in (\ref{prob:minPtildeRef}) still has nonconvex constraints (\ref{cons:minPrad}) and (\ref{cons:minPUL}).

To obtain a more tractable form, we employ the SCA technique to handle constraints (\ref{cons:minPrad}) and (\ref{cons:minPUL}).
Focusing on (\ref{cons:minPrad}), due to the fact that $\{ \mathbf V_{l} \succeq \mathbf 0\}_{l=0}^L$, we have $\mathbf {\bar Q} \succeq \mathbf 0$ and $\mathbf a_t^H(\theta_0) \mathbf {\bar Q} \mathbf a_t (\theta_0) \geq 0$. Moreover, it must hold that $\mathbf a_t^H(\theta_0) \mathbf {\bar Q} \mathbf a_t (\theta_0) \neq 0$ since the radar SINR threshold $\tau^\text{rad} >0$. Thus, by dividing both sides of (\ref{cons:minPrad}) by $\mathbf a_t^H(\theta_0) \mathbf {\bar Q} \mathbf a_t (\theta_0)$, it becomes
\begin{align}\label{cons1}
\mathbf a_r^H(\theta_0)\mathbf \Psi^{-1} \mathbf a_r(\theta_0)
\geq \frac{\tau^\text{rad}}{|\beta_0|^2} \left( \mathbf a_t^H(\theta_0) \mathbf {\bar Q} \mathbf a_t (\theta_0) \right)^{-1}.
\end{align}
Considering that the function $f(\mathbf Y) = \mathbf f^H \mathbf Y^{-1} \mathbf f$ is convex with respect to $\mathbf Y$ for $\mathbf Y \succ \mathbf 0$ \cite[Section 3.1.7]{cvx} and $\mathbf \Psi$ is an affine function of $\{\{\mathbf {V}_l \}_{l=0}^L,\{p_k \}_{k=1}^K \}$, the left-hand side of (\ref{cons1}) is convex with respect to $\mathbf \Psi$, and is also convex with respect to $\{\{\mathbf {V}_l \}_{l=0}^L,\{p_k \}_{k=1}^K \}$ \cite[Section 3.2.2]{cvx}. Similarly, the right-hand side of (\ref{cons1}) is convex with respect to $\{\mathbf {V}_l \}_{l=0}^L$.
Therefore, (\ref{cons1}) is a difference-of-convex (DC) constraint which can be handled by iteratively lower bounding the left-hand side by its first-order Taylor expansion \cite{CCCP}.
Specifically, according to the complex-valued derivatives in \cite{Derivatives}, for the $i$-th iteration of the SCA, we consider the following lower bound:
\begin{align}\label{firstTE:psi}
&\ \mathbf a_r^H(\theta_0)\mathbf \Psi^{-1} \mathbf a_r(\theta_0)\nonumber \\
\geq&\   \mathbf a_r^H(\theta_0)\left(\mathbf \Psi^{(i-1)}\right)^{-1} \mathbf a_r(\theta_0) - \mathbf a_r^H(\theta_0)\left(\mathbf \Psi^{(i-1)}\right)^{-1}  \nonumber\\
&\ \times \left(\mathbf \Psi-\mathbf \Psi^{(i-1)}\right)\left(\mathbf \Psi^{(i-1)}\right)^{-1} \mathbf a_r(\theta_0) \nonumber \\
 \triangleq&\ f\left(\mathbf \Psi, \mathbf \Psi^{(i-1)}\right),
\end{align}
where $\mathbf \Psi^{(i-1)}\triangleq \sum_{k=1}^K p_k^{(i-1)}  \mathbf h_k \mathbf h_k^H +  \mathbf B \mathbf {\bar Q}^{(i-1)} \mathbf B^H + \sigma_r^2 \mathbf I_{N_r}$ and $\mathbf {\bar Q}^{(i-1)} = \sum_{l=0}^L \mathbf V_l^{(i-1)}$ with $\{p_k^{(i-1)} \}_{k=1}^K$ and $\{\mathbf V_l^{(i-1)} \}_{l=0}^L$ being the solutions obtained in the $(i-1)$-th iteration.
As such, a convex subset of the nonconvex constraint in (\ref{cons:minPrad}) is established as
\begin{align}\label{cvxApprox:rad}
f\left(\mathbf \Psi, \mathbf \Psi^{(i-1)}\right) \geq \frac{\tau^\text{rad}}{|\beta_0|^2} \left( \mathbf a_t^H(\theta_0) \mathbf {\bar Q} \mathbf a_t (\theta_0) \right)^{-1}.
\end{align}

Next, we consider the constraints in (\ref{cons:minPUL}). For each $k$, it can be similarly verified that $\mathbf h_k^H  \mathbf \Phi_k^{-1} \mathbf h_k $ is convex with respect to $\mathbf \Phi_k$ and we thus exploit its first-order Taylor expansion to obtain an affine approximation as
\begin{align}\label{firstTE:phi}
 \mathbf h_k^H  \mathbf \Phi_k^{-1} \mathbf h_k
\geq&\  \mathbf h_k^H\left(\mathbf \Phi_k^{(i-1)}\right)^{-1} \mathbf h_k
 - \mathbf h_k^H\left(\mathbf \Phi_k^{(i-1)}\right)^{-1} \nonumber\\
 &\ \times \left(\mathbf \Phi_k-\mathbf \Phi_k^{(i-1)}\right) \left(\mathbf \Phi_k^{(i-1)}\right)^{-1} \mathbf h_k \nonumber \\
\triangleq&\ f_k\left(\mathbf \Phi_k, \mathbf \Phi_k^{(i-1)}\right), \ \forall k,
\end{align}
where $\mathbf \Phi_k^{(i-1)}\triangleq \sum_{k'=1,k'\neq k}^K p_{k'}^{(i-1)}  \mathbf h_{k'} \mathbf h_{k'}^H +  \mathbf C \mathbf {\bar Q}^{(i-1)} \mathbf C^H + \sigma_r^2 \mathbf I_{N_r}$ is calculated based on the solutions obtained in the $(i-1)$-th iteration.
With (\ref{firstTE:phi}), a convex subset of (\ref{cons:minPUL}) is given by
\begin{align}\label{cvxApprox:comk}
 f_k\left(\mathbf \Phi_k, \mathbf \Phi_k^{(i-1)}\right) \geq \frac{\tau_k^\text{com,UL}} {p_k},\ \forall k.
\end{align}

Based on the convex approximations in (\ref{firstTE:psi}) and (\ref{firstTE:phi}), we are ready to obtain a series of surrogate problems to locally approximate (\ref{prob:minPtildeRef}). Specifically, the surrogate problem in the $i$-th iteration is formulated as
\begin{align}\label{prob:minPSCA}
\mathop \text{minimize} \limits_{ \{\mathbf V_l \succeq \mathbf 0 \}_{l=0}^L, \atop \{p_k \geq 0 \}_{k=1}^K } \quad
& \sum_{l=0}^L \text{Tr}(\mathbf V_l) + \sum_{k=1}^K p_k \nonumber \\
\text{subject to}\quad
&\rm (\ref{cons:minPDL}),(\ref{cvxApprox:rad}), (\ref{cvxApprox:comk}).
\end{align}
This problem is convex and its globally optimal solution can be obtained via, e.g., the interior point method \cite{cvx} or some off-the-shelf convex optimization tools, e.g., CVX \cite{CVXtool}.
After solving (\ref{prob:minPSCA}), we update $\mathbf {\bar Q}^{(i)}$, $\mathbf \Psi^{(i)}$, and $\{ \mathbf \Phi_k^{(i)}\}_{k=1}^K$ by exploiting the optimal solutions to $\{\{\mathbf V_l\}_{l=0}^L, \{p_k\}_{k=1}^K\}$ and then proceed to the $(i+1)$-th iteration.
Furthermore, according to \cite{SCAconvergence}, this iterative procedure converges to a Karush-Kuhn-Tucker (KKT) point of the problem in (\ref{prob:minPtildeRef}).

Upon convergence, we denote the obtained solution as $\{\{\mathbf {\widehat V}_l\}_{l=0}^L, \{\hat p_k \}_{k=1}^K\}$. An additional procedure, such as Gaussian randomization \cite{GaussianR}, is generally exploited for recovering the beamforming vectors $\{\mathbf v_l\}_{l=1}^L$, i.e., the solution of (\ref{prob:minPtilde}), since $\{\mathbf {\widehat V}_l\}_{l=1}^L$ may not satisfy the relaxed rank-one constraints in (\ref{cons:rank1}).
However, the commonly adopted Gaussian randomization for recovering a rank-one solution generally has high computational complexity and leads to certain performance loss. Fortunately, based on the following theorem, we prove that a rank-one solution of (\ref{prob:minPtildeRef}) can always be constructed from $\{\mathbf {\widehat V}_l\}_{l=1}^L$ without performance loss.

\begin{theorem}\label{theorem:rank1soltuion}
Based on the solution $(\{\mathbf {\widehat V}_l\}_{l=0}^L, \{\hat p_k \}_{k=1}^K)$ in-hand, a solution of (\ref{prob:minPtildeRef}) achieving the same power consumption as $(\{\mathbf {\widehat V}_l\}_{l=0}^L, \{\hat p_k \}_{k=1}^K)$ while satisfying the relaxed rank-one constraints in (\ref{cons:rank1}) can be constructed as
\begin{align}
\mathbf {V}^*_l =&\ \mathbf {v}^*_l ({\mathbf v}_l^*)^H, \ \forall l\geq 1, \nonumber\\
\mathbf {V}_0^* =&\ \sum_{l=1}^L \mathbf {\widehat V}_l + \mathbf {\widehat V}_0 - \sum_{l=1}^L \mathbf {v}_l^* ({\mathbf {v}^*_l})^H, \nonumber\\
p_k^* =&\ \hat p_k, \ \forall k. \label{def:tildeV0}
\end{align}
where $\mathbf {v}^*_l = \left(\mathbf g_l^H \mathbf {\widehat V}_l \mathbf g_l \right)^{-1/2} \mathbf {\widehat V}_l \mathbf g_l, \ \forall l\geq 1.$
\end{theorem}
\begin{IEEEproof}
See Appendix \ref{proof:theoremrank1soltuion}.
\end{IEEEproof}

\begin{algorithm}[t]
\caption{Proposed Algorithm to Solve (\ref{prob:minP})}
\label{alg:minP}
\begin{algorithmic}[1]
\STATE \textit{Initialization:} Initialize $\{\{\mathbf V_l^{(0)}\}_{l=0}^L, \{p_k^{(0)} \}_{k=1}^K \}$, iteration index $i = 0$, and convergence accuracy $\epsilon$.
\REPEAT
    \STATE Set $i = i + 1$.\\
    \STATE
    Solve (\ref{prob:minPSCA}) with $\{ \{\mathbf V_l^{(i-1)}\}_{l=0}^L, \{p_k^{(i-1)} \}_{k=1}^K \}$ and update $\{ \{\mathbf V_l^{(i)}\}_{l=0}^L, \{p_k^{(i)} \}_{k=1}^K\}$.\\

\UNTIL \textit{Convergence}.
\STATE Calculate the transmit beamforming and the uplink transmit power according to (\ref{def:tildeV0}).
\STATE Calculate the receive beamformers according to (\ref{w_r*}) and (\ref{w_k*}), respectively.
\STATE \textit{Output:} $\mathbf u$, $\{\mathbf w_k \}_{k=1}^K$, $\{\mathbf v_l \}_{l=1}^L$, $\mathbf V_0$, and $\{p_k \}_{k=1}^K$.
\end{algorithmic}
\end{algorithm}

Theorem \ref{theorem:rank1soltuion} indicates that we can obtain a new solution of (\ref{prob:minPtildeRef}) by $\{\{\mathbf {V}^*_l\}_{l=0}^L, \{p_k^* \}_{k=1}^K\}$, which
satisfies the rank-one constraints in (\ref{cons:rank1}) and attains the same performance as a KKT solution, i.e., $\{\{\mathbf {\widehat V}_l\}_{l=0}^L, \{\hat p_k \}_{k=1}^K\}$. At the same time, we recover the solution of (\ref{prob:minPtilde}) as $\{\{\mathbf {v}_l^*\}_{l=1}^L, \mathbf {V}_0^*,\{p_k^*\}_{k=1}^K\}$. The corresponding receive beamformers are further calculated based on (\ref{w_r*}) and (\ref{w_k*}).
We summarize the procedure for solving the power minimization problem in (\ref{prob:minP}) as Algorithm~\ref{alg:minP}.
Note that the main computational burden of Algorithm \ref{alg:minP} stems from solving (\ref{prob:minPSCA}) in each iteration.
Following \cite[Section V-A]{complexity}, which presents a detailed method to quantitatively analyze the computational complexity of solving a convex problem through the interior point method, we obtain the complexity order for solving (\ref{prob:minPSCA}) as $\mathcal O\left(\sqrt{N_tL+K}( N_t^{6} L^{3} + N_t^{4} L^{2} K + K^{3}) \right)$.

\subsection{Special Case of Uplink Communication Only}
The above considered problem involves FD communication. We herein focus on a special case in the absence of downlink communication, i.e., $L=0$, and the downlink signal is used for target detection only, which is similar to the scenario integrating sensing with uplink communication in \cite{M.Temiz2021TCCN,X.Wang2022CL}. The difference is that a fixed communication signal is considered in \cite{M.Temiz2021TCCN,X.Wang2022CL} while we also optimize the uplink transmission here.
In this special case, the downlink transmit signal $\mathbf x$ and its covariance matrix $\mathbf Q$ reduce to
\begin{align}
 \mathbf x = \mathbf s_0, \quad
 \mathbf Q = \mathbf V_0.
\end{align}
As a result, problem (\ref{prob:minP}) becomes
\begin{align}\label{prob:minP_sensingonly}
\mathop \text{minimize} \limits_{\mathbf u, \{ \mathbf w_k\}_{k=1}^K,\atop \mathbf V_0 \succeq \mathbf 0, \{p_k \geq0\}_{k=1}^K}
& \text{Tr}(\mathbf V_0) + \sum_{k=1}^K p_k \nonumber \\
\text{subject to}\quad
& \gamma^\text{rad} \geq \tau^\text{rad},
\ \gamma_k^\text{com,UL} \geq \tau_k^\text{com,UL}, \ \forall k.
\end{align}
Compared to (\ref{prob:minP}), this problem removes the optimizations on downlink communication  and the proposed SCA-based Algorithm~\ref{alg:minP} is also applicable. However, to solve this simplified problem, we are able to develop an alternative algorithm which dramatically reduces the computational complexity.

Different from Algorithm \ref{alg:minP}, we consider solving (\ref{prob:minP_sensingonly}) by optimizing $\left\{\mathbf u, \{ \mathbf w_k\}_{k=1}^K \right\}$ and $\{\mathbf V_0, \{p_k\}_{k=1}^K\}$ in an alternating manner. Recalling that given $\{\mathbf V_0, \{p_k\}_{k=1}^K\}$, the optimal solutions to $\left\{\mathbf u, \{ \mathbf w_k\}_{k=1}^K \right\}$ are obtained in closed-form expressions as shown in Proposition \ref{prop:receiver}, it remains to optimize $\{\mathbf V_0, \{p_k\}_{k=1}^K\}$ with fixed $\left\{\mathbf u, \{ \mathbf w_k\}_{k=1}^K \right\}$.
Define the following constant terms ${\tilde a}_k \triangleq \mathbf w_k^H \mathbf h_k \mathbf h_k^H \mathbf w_k$,
$\mathbf {\tilde b}_k \triangleq \mathbf C^H \mathbf w_k$,
${\tilde c}_{k,k'} \triangleq \mathbf w_k^H \mathbf h_{k'} \mathbf h_{k'}^H \mathbf w_k, \forall k' \neq k$,
and $ {\tilde d}_k\triangleq \sigma_r^2 \mathbf w_k^H \mathbf w_k$ to simplify the formulations of $\gamma_k^\text{com,UL},\ \forall k,$ in (\ref{sinr_k}).
Also, define $\mathbf {\tilde e} \triangleq | \beta_0|\mathbf A^H(\theta_0) \mathbf u$,
$\mathbf {\tilde f} \triangleq \mathbf B^H \mathbf u$,
${\tilde g}_k \triangleq \mathbf u^H \mathbf h_{k} \mathbf h_{k}^H \mathbf u,\ \forall k,$
and ${\tilde h} \triangleq  \sigma_r^2 \mathbf u^H \mathbf u$ to simplify $\gamma^\text{rad}$ in (\ref{sinr_r}). Then, the subproblem with respect to $\{\mathbf V_0, \{p_k\}_{k=1}^K\}$ is expressed as
\begin{align}\label{prob:minP_SCSDP}
\mathop \text{minimize} \limits_{\mathbf V_0 \succeq \mathbf 0,\atop \{p_k \geq0\}_{k=1}^K} \quad
& \text{Tr}(\mathbf V_0) + \sum_{k=1}^K p_k \nonumber \\
\text{subject to}\quad
& \frac{\mathbf {\tilde e}^H \mathbf V_0 \mathbf {\tilde e}}
{\mathbf {\tilde f}^H \mathbf V_0 \mathbf {\tilde f} + \sum_{k}  {\tilde g}_kp_k + {\tilde h}} \geq \tau^\text{rad}, \nonumber \\
& \frac{ {\tilde a}_k p_k}{ \mathbf {\tilde b}_k^H \mathbf V_0 \mathbf {\tilde b}_k + \sum_{k'\neq k}{\tilde c}_{k,k'}p_{k'} + {\tilde d}_k } \geq \tau_k^\text{com,UL},  \ \forall k.
\end{align}
By rearranging the constraints, this problem is readily transformed into a standard semidefinite programming (SDP) and solved directly. Nonetheless, we will show that (\ref{prob:minP_SCSDP}) is equivalent to an SOCP, which can be solved with much lower computational complexity compared to that of the SDP formulation.

To begin with, we prove the following result regarding (\ref{prob:minP_SCSDP}).

\begin{theorem}\label{theorem:V0rank}
The optimal solution to (\ref{prob:minP_SCSDP}), denoted by $\{\mathbf V_0^*,\{p_k^*\}_{k=1}^K\}$, must satisfy
\begin{align}
\text{rank}(\mathbf V_0^*) = 1.
\end{align}
\end{theorem}
\begin{IEEEproof}
See Appendix \ref{proof:theoremV0rank}.
\end{IEEEproof}
With Theorem \ref{theorem:V0rank}, we further obtain the following proposition.

\begin{proposition}\label{prop:minPSOCP}
Let us introduce $\mathbf v_0 \in \mathbb C^{N_t \times 1}$ and real-valued $t_0$ and $\{ q_k\}_{k=1}^K$ as optimization variables. The optimal solution of (\ref{prob:minP_SCSDP}) can be achieved by solving the following SOCP
\begin{align}\label{prob:minP_sensingonlySOCP}
\mathop \text{minimize} \limits_{ \mathbf v_0,t_0,\{ q_k\}_{k=1}^K } \quad
& t_0 \nonumber \\
\text{subject to}\quad
& \left[ \begin{matrix} \mathbf {\tilde e}^H \mathbf v_0 \\ \boldsymbol \varpi \end{matrix} \right] \succeq_C \mathbf 0,
\ \left[ \begin{matrix} \sqrt{{\tilde a}_k} q_k \\ \boldsymbol \varrho_k \end{matrix} \right] \succeq_C \mathbf 0, \ \forall k, \nonumber\\
& \left[ \begin{matrix} t_0 \\ \mathbf v_0 \\ \mathbf q \end{matrix} \right] \succeq_C \mathbf 0,
\end{align}
where $\boldsymbol \varpi \triangleq \sqrt{\tau^\text{rad}} [ \mathbf {\tilde f}^H \mathbf v_0, \sqrt{{\tilde g}_1} q_1, \cdots, \sqrt{{\tilde g}_K} q_K, \sqrt{{\tilde h}}]^T$,
$\boldsymbol \varrho_k \triangleq \sqrt{\tau_k^\text{com,UL}}[\mathbf {\tilde b}_k^H \mathbf v_0, \sqrt{ {\tilde c}_{k,1}} q_{1},\cdots, \sqrt{ {\tilde c}_{k,k-1}}q_{k-1}, \sqrt{ {\tilde c}_{k,k+1}}q_{k+1},\\ \cdots, \sqrt{ {\tilde c}_{k,K}}q_{K}, \sqrt{{\tilde d}_k} ]^T, \ \forall k$, and $\mathbf q \triangleq [q_1, \cdots, q_K]^T$.
The notation $\succeq_C$ denotes the generalized inequality as
$\left[ \begin{matrix} z \\ \mathbf z \end{matrix} \right] \succeq_C\mathbf 0 \iff \|\mathbf z \| \leq z$\cite{A.Wiesel2006Linear}.
Denoting the optimal solution of (\ref{prob:minP_sensingonlySOCP}) as $\{\mathbf v_0^*,t_0^*,\{ q_k^*\}_{k=1}^K\}$, we can derive the optimal solution of (\ref{prob:minP_SCSDP}) by $\mathbf V_0^* = \mathbf v_0^* (\mathbf v_0^*)^H$ and $p_k^* = (q_k^*)^2,\ \forall k.$
\end{proposition}
\begin{IEEEproof}
See Appendix~\ref{proof:propminPSOCP}.
\end{IEEEproof}
Proposition \ref{prop:minPSOCP} provides a computationally efficient approach to find the optimal solution of subproblem (\ref{prob:minP_SCSDP}).

\begin{algorithm}[t]
\caption{Low-Complexity Solution for (\ref{prob:minP_sensingonly})}
\label{alg:minPSC}
\begin{algorithmic}[1]
\STATE \textit{Initialization:} Initialize $\{\mathbf V_0^{(0)},\{ p_k^{(0)}\}_{k=1}^K \}$. Set iteration index $i = 0$ and convergence accuracy $\epsilon$.
\REPEAT
    \STATE Set $i = i + 1$.\\
    \STATE Update $\{ \mathbf u^{(i)},\{\mathbf w_k^{(i)} \}_{k=1}^K \}$ with $\{\mathbf V_0^{(i-1)}, \{ p_k^{(i-1)}\}_{k=1}^K\}$ according to (\ref{w_r*}) and (\ref{w_k*}), respectively.
    \STATE
     Solve the SOCP in (\ref{prob:minP_sensingonlySOCP}) with $\{\mathbf u^{(i)},\{\mathbf w_k^{(i)} \}_{k=1}^K \}$ and update $\{\mathbf V_0^{(i)}, \{ p_k^{(i)}\}_{k=1}^K\}$.\\

\UNTIL \textit{Convergence.}
\STATE \textit{Output:} $\mathbf u$, $\{\mathbf w_k \}_{k=1}^K$, $\mathbf V_0$, and $\{ p_k\}_{k=1}^K$.
\end{algorithmic}
\end{algorithm}

Finally, the proposed low-complexity method for solving (\ref{prob:minP_sensingonly}) is summarized in Algorithm~\ref{alg:minPSC}. It is easily verified that the objective value of the power consumption is nonincreasing over the iterations and the solution set is compact, thus the
proposed algorithm is guaranteed to converge.
In terms of computational complexity,
in each round of the iteration, the dominating computations of updating $\mathbf u^{(i)}$ and $\{\mathbf w_k^{(i)} \}_{k=1}^K$ lie in the calculation of matrix inversion, leading to the complexity of $\mathcal O \left(K N_r^3\right)$.
For updating $\mathbf V_0^{(i)}$ and $\{p_k^{(i)}\}_{k=1}^K$, according to the complexity analysis in \cite[Section V-A]{complexity}, the complexity order of solving the SOCP in (\ref{prob:minP_sensingonlySOCP}) is given by $\mathcal O\left( \sqrt{K} (K^{4} + N_t^3 + N_tK^{3}) \right)$.
For comparison, we restate the computational complexity of Algorithm~\ref{alg:minP} in this special case as $\mathcal O\left(\sqrt{N_t+K}( N_t^{6} + N_t^{4} K + K^{3}) \right)$.
It is found that Algorithm~\ref{alg:minPSC} enjoys a much lower order of computational cost than that of Algorithm~\ref{alg:minP} in each iteration.
Furthermore, as will be shown in Section~V, these two algorithms share similar convergence speed and performance. Hence, for this special case, Algorithm~\ref{alg:minPSC} is an effective alternative to Algorithm~\ref{alg:minP} with much lower overall complexity.

\section{Joint FD ISAC Design for \\Sum Rate Maximization}
In this section, we consider solving (\ref{prob:maxR}).
Compared to the power minimization criterion, (\ref{prob:maxR}) focuses on the sum rate maximization which is generally NP-hard, even in communication-only systems \cite{LuoJSTSP2008Dynamic,C.XingTSP2020Newpoint}.
To solve this difficult problem, we first predetermine the optimal receivers and then develop an effective iterative algorithm.

\subsection{Problem Reformulation}
Similarly to the previous section, by predetermining the optimal receive beamformers in (\ref{w_r*}) and (\ref{w_k*}) and invoking the equivalent SINR expressions in (\ref{sinr_r'}) and (\ref{sinr_k'}), we rewrite (\ref{prob:maxR}) into the following form:
\begin{align}\label{prob:maxRnew}
\mathop \text{maximize} \limits_{\{\mathbf v_l\}_{l=1}^L,\mathbf V_0 \succeq \mathbf 0,\atop \{p_k \geq 0\}_{k=1}^K} \
& \sum_{k=1}^K \log_2(1 + \bar{\gamma}_k^\text{com,UL}) + \sum_{l=1}^L \log_2(1 + \gamma_l^\text{com,DL}) \nonumber \\
\text{subject to}\quad
& \bar{\gamma}^\text{rad} \geq \tau^\text{rad},  \nonumber \\
& \sum_{l=1}^L \| \mathbf v_l \|^2 + \text{Tr}(\mathbf V_0) \leq P_\text{max}, \nonumber \\
& p_k \leq P_k, \ \forall k.
\end{align}
By introducing a group of real-valued auxiliary optimization variables $u_k \geq 0, \ \forall k,$ and defining $\mathbf V_l \triangleq \mathbf v_l \mathbf v_l^H, \ \forall l$, we further reformulate (\ref{prob:maxRnew}) as
\begin{align}\label{prob:maxRRef}
\mathop \text{maximize}\limits_{\{ \mathbf V_l \succeq \mathbf 0\}_{l=0}^L, \atop \{p_k\geq0,u_k \geq0\}_{k=1}^K} & \sum_{k=1}^K \log_2(1+u_k) + \sum_{l=1}^L \log_2(1+ \bar{\gamma}_l^\text{com,DL}) \nonumber \\
\text{subject to} \quad
& \bar{\gamma}^\text{rad} \geq \tau^\text{rad}, \ \text{Tr}\left( \sum_{l=0}^L \mathbf V_l\right) \leq P_\text{max}, \nonumber \\
& u_k \leq \bar{\gamma}_k^\text{com,UL},\ p_k \leq P_k,\ \forall k,
\end{align}
where
$\bar{\gamma}_l^\text{com,DL} = \frac{\mathbf g_l^H \mathbf V_l \mathbf g_l}{\sum_{l'=1,l'\neq l}^L \mathbf g_l^H \mathbf V_{l'} \mathbf g_l + \mathbf g_l^H \mathbf V_0 \mathbf g_l  + \sigma^2_l}$
and the rank-one constraints of $\{ \mathbf V_l\}_{l=1}^L$ are omitted here.
It is easily proved by contradiction that the constraints $u_k \leq \bar{\gamma}_k^\text{com,UL}, \ \forall k,$ must keep active at the optimality, which verifies the equivalence between problems (\ref{prob:maxRnew}) and (\ref{prob:maxRRef}).

\subsection{Proposed Solution}
For solving (\ref{prob:maxRRef}), the difficulties lie in the nonconcave term $\sum_{l=1}^L \log_2(1+ \bar{\gamma}_l^\text{com,DL})$ in the objective function and the nonconvex constraints $\bar{\gamma}^\text{rad} \geq \tau^\text{rad}$ and $ u_k \leq \bar{\gamma}_k^\text{com,UL},\ \forall k$.
To tackle the nonconvexity of the objective function, we first rewrite the achievable rate of downlink user $l$ as
\begin{align}\label{log2DL}
\log_2(1\!+ \bar{\gamma}_l^\text{com,DL}) =&\ \log_2 \! \left( \sum_{l'=0}^L \mathbf g_l^H \mathbf V_{l'} \mathbf g_l  \!+ \! \sigma^2_l\right) \nonumber \\
&\ -\! \log_2\!\left(\sum_{l'=0,l'\neq l}^L \mathbf g_l^H \mathbf V_{l'} \mathbf g_l \!+ \!\sigma^2_l\right).
\end{align}
Both logarithm functions are concave with respect to $\{\mathbf V_l\}_{l=0}^L$ and we thus can approximate (\ref{log2DL}) via linearizing the second term via the SCA approach. Specifically, by exploiting the first-order Taylor expansion, it holds that
\begin{align}\label{taylorrate_l}
&\ \log_2\left(\sum_{l'=0,l'\neq l}^L \mathbf g_l^H \mathbf V_{l'} \mathbf g_l + \sigma^2_l\right) \nonumber \\
\leq&\  a_l^{(j-1)} + \frac{\log_2 \rm e}{2^{a_l^{(j-1)}}} \sum_{l'=0,l'\neq l}^L \mathbf g_l^H  \left(\mathbf V_{l'} - \mathbf V_{l'}^{(j-1)} \right) \mathbf g_l  \nonumber \\
\triangleq&\ \underline{r}_l,
\end{align}
where $\mathbf V_{l'}^{(j-1)}$ is the solution to $\mathbf V_{l'}$ obtained in the $(j-1)$-th iteration and $a_l^{(j-1)} = \log_2\left( \sum_{l'=0,l'\neq l}^L \mathbf g_l^H  \mathbf V_{l'}^{(j-1)} \mathbf g_l + \sigma_l^2 \right)$. Applying (\ref{taylorrate_l}), we establish a lower bound of $\log_2(1+ \bar{\gamma}_l^\text{com,DL})$ as follows
\begin{align}
\log_2(1+ \bar{\gamma}_l^\text{com,DL}) \geq \log_2\left( \sum_{l'=0}^L \mathbf g_l^H \mathbf V_{l'} \mathbf g_l  + \sigma^2_l\right) - \underline{r}_l,
\end{align}
which is concave with respect to $\{ \mathbf V_{l} \}_{l=0}^L$.

Next, we handle the nonconvex constraints.
Note that the approximations utilized in Algorithm \ref{alg:minP} can still be applied here to deal with these constraints. Specifically, the radar SINR constraint can be tackled as (\ref{cvxApprox:rad}). Applying (\ref{firstTE:phi}) to $ u_k \leq \bar{\gamma}_k^\text{com,UL},\ \forall k,$ yields
\begin{align}\label{cons:u/p}
\frac{u_k}{p_k} \leq f_k\left(\mathbf \Phi_k, \mathbf \Phi^{(j-1)}_k\right),\ \forall k.
\end{align}
However, different from (\ref{cvxApprox:comk}), the fractional function $\frac{u_k}{p_k}$ in the left-hand side makes constraint (\ref{cons:u/p}) still nonconvex. In order to handle this issue, by introducing real-valued auxiliary variables $\{ x_k \}_{k=1}^K$, we first equivalently transform the single constraint in (\ref{cons:u/p}) into two separate constraints as follows
\begin{align}\label{cons:x2}
\begin{cases}
\frac{x_k^2}{p_k} \leq f_k\left(\mathbf \Phi_k, \mathbf \Phi^{(j-1)}_k\right), \\
\ u_k \leq x_k^2,
\end{cases}
\ \forall k,
\end{align}
where the function $\frac{x_k^2}{p_k}$ is convex when $p_k > 0$ \cite{cvx}.
This equivalence is established based on the fact that the constraint $\ u_k \leq x_k^2$ must keep active at the optimality, which can be proved by contradiction. Then, by replacing the convex function $x_k^2$ by its first Taylor expansion, we obtain a convex approximation of (\ref{cons:x2}) as
\begin{align}\label{cons:x}
\begin{cases}
\frac{x_k^2}{p_k} \leq f_k\left(\mathbf \Phi_k, \mathbf \Phi^{(j-1)}_k\right), \\
\ u_k \leq \left(x_k^{(j-1)} \right)^2 + 2x_k^{(j-1)} \left(x_k - x_k^{(j-1)} \right),
\end{cases}
\ \forall k,
\end{align}
where the variables $\{x_k^{(j-1)}\}_{k=1}^K$ are iteratively updated in each iteration.

Now, (\ref{prob:maxRRef}) can be addressed by employing the SCA framework. In particular, in the $j$-th iteration we consider the following convex optimization problem
\begin{align}\label{prob:ISAC_convex}
\mathop \text{maximize}\limits_{\{ \mathbf V_l \succeq \mathbf 0 \}_{l=0}^L, \atop \{p_k \geq 0,u_k \geq 0,x_k \geq 0\}_{k=1}^K }  & \sum_{k=1}^K \log_2(1+u_k) \nonumber\\
& + \sum_{l=1}^L\left( \log_2\left( \sum_{l'=0}^L \mathbf g_l^H \mathbf V_{l'} \mathbf g_l  + \sigma^2_l\right) - \underline{r}_l \right) \nonumber \\
\text{subject to} \quad\ \
&\text{Tr}\left( \sum_{l=0}^L \mathbf V_l \right) \leq P_\text{max}, \  p_k \leq P_k,\ \forall k, \nonumber\\
& (\ref{cvxApprox:rad}),\ (\ref{cons:x}),
\end{align}
whose globally optimal solution can thus be readily found. After solving (\ref{prob:ISAC_convex}), we update $\mathbf {\bar Q}^{(j)}$, $\mathbf \Psi^{(j)}$, and $\{\mathbf \Phi^{(j)}_k, x_k^{(j)} \}_{k=1}^K$.
By iteratively solving (\ref{prob:ISAC_convex}) until convergence, a KKT point of (\ref{prob:maxRRef}) is obtained \cite{SCAconvergence}.

Moreover, based on this KKT solution, we can similarly construct a rank-one $\{\mathbf V_l^* \}_{l=1}^L$ without any performance loss according to Theorem~\ref{theorem:rank1soltuion}. The proof procedure is similar to Appendix~\ref{proof:theoremrank1soltuion} and omitted.
Then, the receive beamformers can be calculated according to Proposition \ref{prop:receiver}.
The proposed algorithm for solving (\ref{prob:maxR}) follows a similar procedure as Algorithm \ref{alg:minP} except that in step 4) problem (\ref{prob:minPSCA}) is replaced by problem (\ref{prob:ISAC_convex}) and we need update $\{\mathbf V_l^{(j)} \}_{l=0}^L$ and $\{ p_k^{(j)},x_k^{(j)} \}_{k=1}^K$. We denote it as Algorithm 3 and no longer present the details for brevity. Moreover, the main computational cost of Algorithm~3 lies in solving (\ref{prob:ISAC_convex}) in each iteration, whose complexity is $\mathcal O\left(\sqrt{N_tL+K}( N_t^{6} L^{3} + N_t^{4} L^{2} K + K^{3}) \right)$.

\section{Simulation Results}
\subsection{Parameter Setup}
Assume that both the transmit and receive ULAs of the BS have $N_t = N_r = 8$ antennas. The BS serves $K=2$ uplink users and $L=2$ downlink users.
For radar sensing, it is assumed that the target of interest is located at $\theta_0 = 0^{\circ}$ and $I=2$ interferers are located at $\theta_1 = -50^{\circ}$ and $\theta_2 = 20^{\circ}$, respectively.
The noise powers at the BS and each downlink user are set to $\sigma^2_r=\sigma^2_l= -70$~dBm, $\forall l$.
We set the maximal transmit power budgets at the BS and each uplink user to $P_\text{max} = 18$~dBW and $P_k = 5$~dBW, $\forall k$, respectively.
Assume that all the user channels follow the LoS channel model \cite{O.E.Ayach2014TWC}, where $\mathbf h_k = \sqrt{\xi_k} \sqrt{N_r} \mathbf a_r (\theta_k^\text{UL}),\ \forall k,$ with $\xi_k$ being the path loss and $\theta_k^\text{UL}$ denoting the user angle direction, and similarly $\mathbf g_l = \sqrt{\xi_l} \sqrt{N_t} \mathbf a_t (\theta_l^\text{DL}),\ \forall l.$
For simplicity, a path loss of -103.6~dB is assumed between each user and the BS.
The directions of downlink users $\{\theta_1^\text{DL},\theta_2^\text{DL}\}$ and uplink users $\{\theta_1^\text{UL},\theta_2^\text{UL}\}$ are set to $\{-40^{\circ}, 60^{\circ}\}$ and $\{45^{\circ}, -75^{\circ}\}$, respectively.
Moreover, the channel power gains of the target and the two interferers are set to $|\beta_0|^2/\sigma^2_r = -30$~dB and ${| \beta_1|^2}/\sigma^2_r={ | \beta_2|^2}/\sigma^2_r = 20$~dB \cite{C.Tsinos2021JSTSP,Wu2018Transmit}.
For the residual SI channel at the BS, we follow \cite{M.Temiz2021TCCN,X.Chen2018Access} and model each entry of $\mathbf H_\text{SI} \in \mathbb C^{N_r \times N_t}$ as $[\mathbf H_\text{SI}]_{p,q} = \sqrt{\alpha^\text{SI}_{p,q}} e^{-j2\pi \frac{d_{p,q}}{\lambda}}$, where $\alpha^\text{SI}_{p,q} > 0$ and $d_{p,q} >0$ denote the residual SI channel power and the distance between the $q$-th transmit antenna and the $p$-th receive antenna, respectively. For simplicity, we set $\alpha^\text{SI} = \alpha^\text{SI}_{p,q} = -110$~dB and let $e^{-j2\pi \frac{d_{p,q}}{\lambda}}$ be a unit-modulus variable with random phase for all the transceiver antenna pairs $(p,q)$.
The required SINR thresholds of downlink communications, uplink communications, and target detection are set to $\tau^\text{com,DL}_l = 12$ dB, $\forall l$, $ \tau^\text{com,UL}_k = 10$ dB, $\forall k$, and $\tau^\text{rad} = 15$~dB, respectively.
All the numerical results are averaged over 200 independent channel realizations.

\subsection{Benchmark Schemes}
For performance comparisons, we introduce the following three benchmark schemes.
\subsubsection{HD Communication-based ISAC}
To show more explicitly the advantages of FD, we consider a benchmark time-division duplex (TDD) transmission scheme (noted as ``HD mode'' in the figures), where the downlink communication and the uplink communication separately occupy two slots while the sensing is continuously performed at the BS for achieving better radar performance (see Appendix~\ref{appendix:HD} for details). The optimization problems in this scheme can be solved using the methods in \cite{Liu2022CRB,C.Tsinos2021JSTSP,J.Pritzker2022Arxiv,L.Chen2022JSAC,M.Temiz2021TCCN,X.Wang2022CL} with some modifications and extensions.
\subsubsection{Communication-only Transmission Scheme}
The second benchmark scheme considers an FD communication-oriented system (noted as ``communication-only'' in the figures) by omitting the sensing SINR constraint. This scheme helps evaluate the impact of integrating the sensing functionality on the communication performance. The optimization problems of this FD communication system can be addressed using the algorithm in~\cite{FDcommun}.
\subsubsection{Sensing-only Scheme}
The third benchmark scheme considers a sensing-oriented system (noted as ``sensing-only'' in the figures) in the absence of communications, where a radar signal $\mathbf s_0$ is sent and the echo is processed by the receiver $\mathbf u$ \cite{Wu2018Transmit,radarCM}. Specifically, for the first
criterion, this scheme minimizes the transmit power of $\mathbf s_0$ while ensuring the radar SINR requirement. For
the second criterion, the scheme maximizes the radar SINR under the transmit power constraint. This scheme shows the impact brought by the integrated communications on radar sensing. The corresponding joint transceiver design can be solved via the algorithms proposed in~\cite{Wu2018Transmit,radarCM}.

Note that the design problems of the three benchmark schemes can also be handled by using the algorithms proposed in this work with some minor modifications.

\subsection{Simulation Results}

\begin{figure}[t]
\begin{center}
      \epsfxsize=7.0in\includegraphics[scale=0.6]{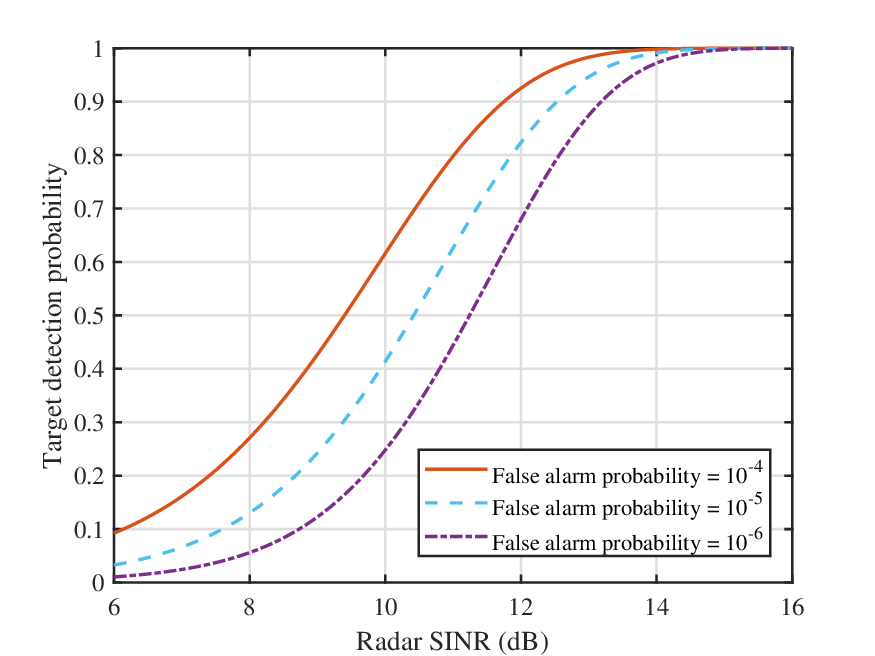}
      \caption{Target detection probability versus radar SINR.}\label{fig:P_detection}
    \end{center}
\end{figure}

In this subsection, the performance of the proposed algorithms is evaluated.
As exemplified in \cite{A.Maiod2008Code}, the authors provided an analytical expression of the target detection probability for the generalized likelihood ratio test detector in a radar system. Invoking the results in \cite[Eq. (4) \& Eq. (6)]{A.Maiod2008Code}, we evaluate the target detection probability with different radar SINRs and false alarm probabilities in Fig. \ref{fig:P_detection}. It is found that given a desired value of the false alarm probability, the detection performance depends on the radar SINR, and a higher detection probability is achieved with the growth of SINR. Thereby, the sensing performance in terms of the detection probability of the system is ensured by setting a relatively large sensing SINR requirement.

\begin{figure}[t]
\begin{center}
      \epsfxsize=7.0in\includegraphics[scale=0.6]{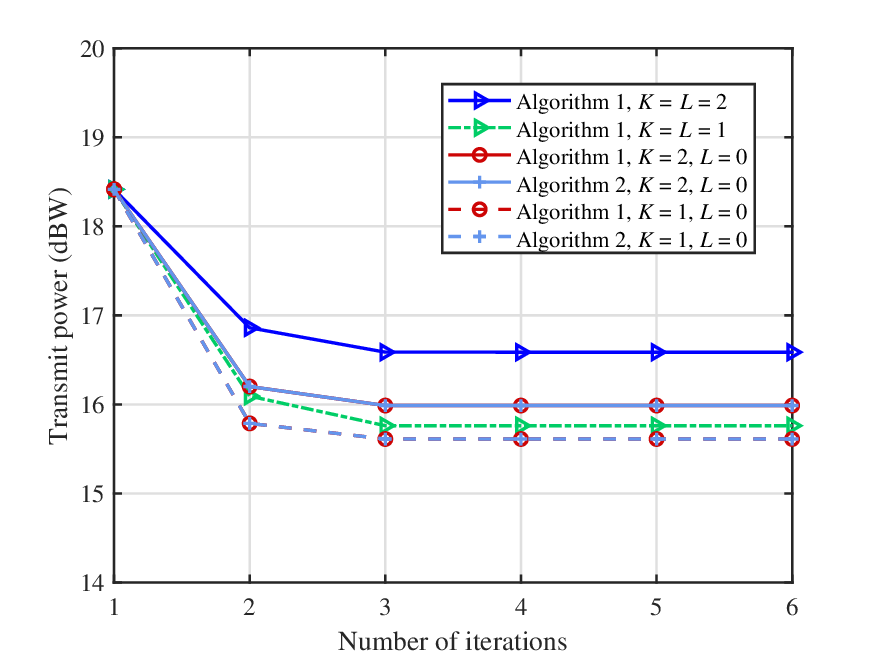}
      \caption{Convergence performances of Algorithm \ref{alg:minP} and Algorithm \ref{alg:minPSC}.}\label{fig:iteration}
    \end{center}
\end{figure}

Fig. \ref{fig:iteration} illustrates the convergence performances of Algorithm~\ref{alg:minP} and Algorithm~\ref{alg:minPSC}.
First, it is found that with different numbers of users, both the algorithms typically converge within 6 iterations. Moreover, it is seen that the power consumption increases with the growth of the number of users, since more transmit power is needed to guarantee the more stringent communication requirements.
On the other hand, concerning the curves with $L$ = 0 that correspond to the special case discussed in Section III-C, it is seen that Algorithm~\ref{alg:minP} and Algorithm~\ref{alg:minPSC} share the same convergence speed and objective value. Recalling that Algorithm~\ref{alg:minPSC} admits a lower computational complexity per iteration than that of Algorithm~\ref{alg:minP}, the complexity advantage of Algorithm~\ref{alg:minPSC} is verified.

\begin{figure}[t]
\begin{center}
      \epsfxsize=7.0in\includegraphics[scale=0.6]{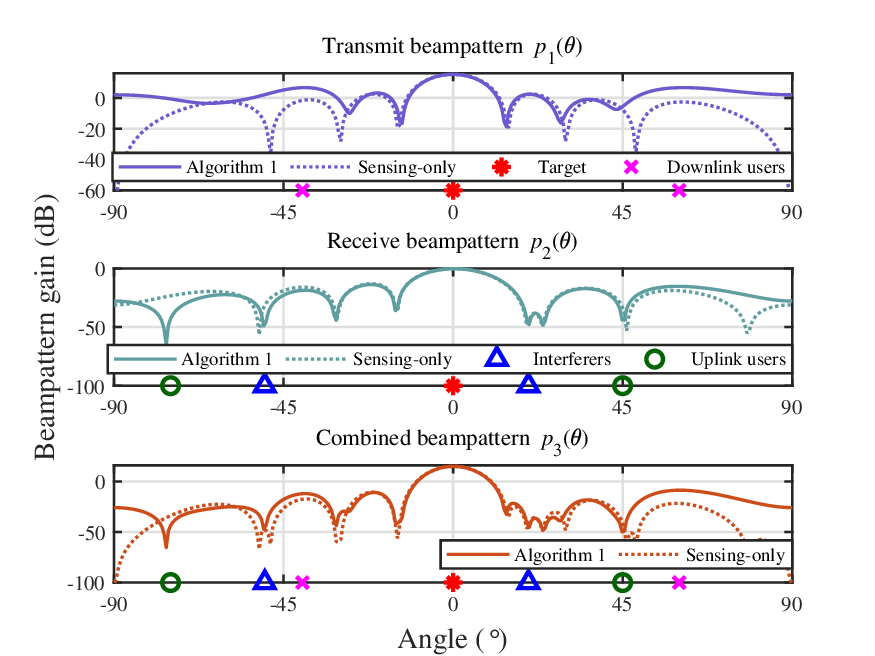}
      \caption{Beampattern regarding radar functionality of Algorithm~\ref{alg:minP}.}\label{fig:radarpattern_of_Alg1}
    \end{center}
\end{figure}

In what follows, we show the beampattern gain regarding the radar functionality achieved by Algorithm~\ref{alg:minP}. Based on the optimized radar receive beamformer $\mathbf u^*$, which has been normalized as $\| \mathbf u^*\|=1$, and the transmit signal $\mathbf x^*$, we define the following beampatterns
\begin{align}
p_1 (\theta) =&\ | \mathbf a_t^H (\theta) \mathbf x^*|^2, \\
p_2 (\theta) =&\ |(\mathbf u^*)^H \mathbf a_r (\theta)|^2 , \\
p_3 (\theta) =&\  | (\mathbf u^*)^H \mathbf a_r (\theta) \mathbf a_t^H (\theta) \mathbf x^*|^2, \label{def:radarpattern}
\end{align}
which show the gain achieved by $\mathbf x^*$, the gain by $\mathbf u^*$, and the joint impact of $\{\mathbf x^*,\mathbf u^*\}$, respectively.
Fig. \ref{fig:radarpattern_of_Alg1} illustrates the above three beampatterns achieved by Algorithm~\ref{alg:minP} and the sensing-only scheme. From the first subfigure, it is seen that three main transmit beams of Algorithm~\ref{alg:minP} are pointed towards the target and the downlink users, respectively.
The second subfigure shows that the directions of the interferers and uplink users are placed with
relatively deep nulls in Algorithm~\ref{alg:minP} since their reflected and transmitted signals do cause severe interference to the radar sensing. The overall beampattern in the third subfigure combines the transmit and receive beampatterns.
Compared with the sensing-only scheme, Algorithm~\ref{alg:minP} can additionally fulfill the requirement of downlink users and suppress the interference stemming from uplink users.
The effectiveness of Algorithm~\ref{alg:minP} regarding the radar functionality is thus validated.

\begin{figure}[t]
\begin{center}
      \epsfxsize=7.0in\includegraphics[scale=0.6]{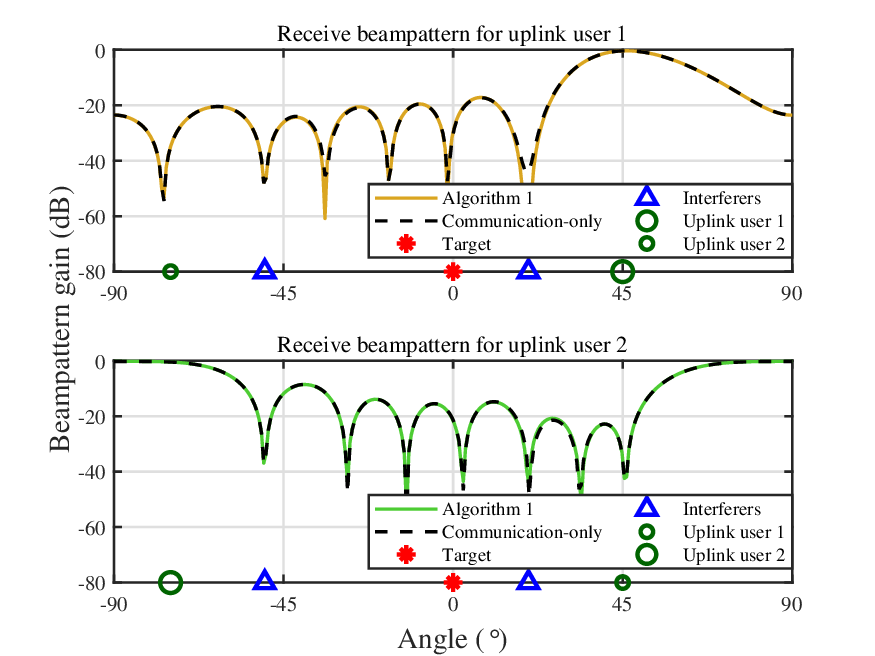}
      \caption{Beampattern regarding communication functionality of Algorithm~\ref{alg:minP}.}\label{fig:communpattern_of_Alg1}
    \end{center}
\end{figure}

Next, we show the beampattern for the communication functionality. With the optimized communication receive beamformers $\{ \mathbf w_k^*\}$, we define the receive beampattern for uplink user $k$ as $| (\mathbf w_k^*)^H \mathbf a_r (\theta)|^2$. The receive beampattern gains for two uplink users are then depicted in Fig. \ref{fig:communpattern_of_Alg1}.
It is seen from the figure that for uplink user 1, $\mathbf w_1^*$ allocates a main beam pointing towards the user direction. Meanwhile, several deep nulls are placed towards the target, interferers, and the other uplink user 2, all of which cause interference when decoding the signal from user 1. Similar observations can be found in the second subfigure for user 2. In addition, our design achieves almost the same pattern as the communication-only scheme. Together with the fact that two main beams of the transmit signal are pointed to the downlink users as shown in the first subfigure in Fig.~\ref{fig:radarpattern_of_Alg1}, the effectiveness of the proposed design in terms of the communication functionality is thus validated.

\begin{figure}[t]
\begin{center}
      \epsfxsize=7.0in\includegraphics[scale=0.6]{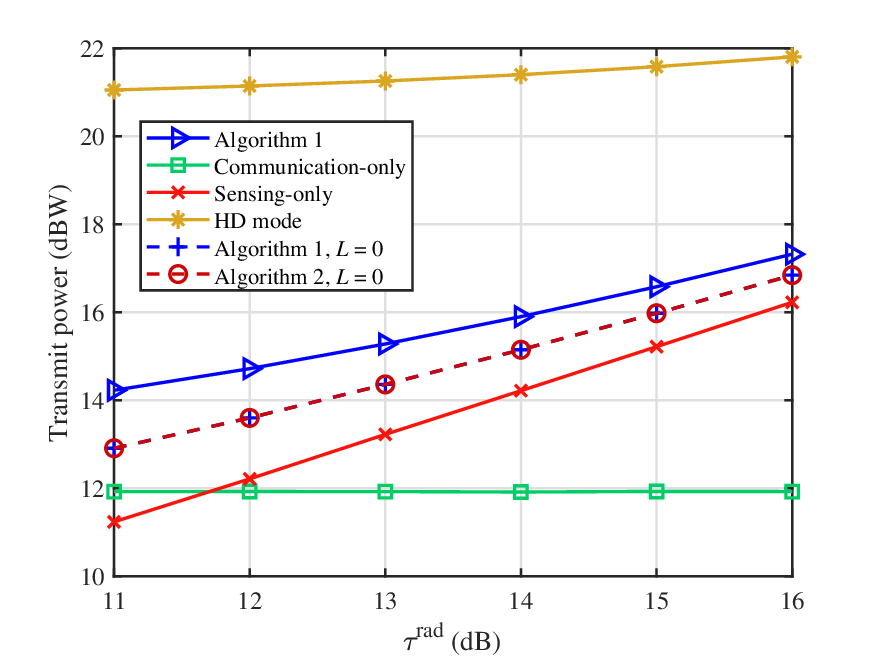}
      \caption{Power consumption versus the radar SINR threshold $\tau^\text{rad}$.}\label{fig:minPversustau}
    \end{center}
\end{figure}

Fig. \ref{fig:minPversustau} shows the minimum total power versus the radar SINR threshold $\tau^\text{rad}$. Observe that the power consumed by the communication-only design remains unchanged since it does not contain the sensing constraint. On the contrary, when $\tau^\text{rad}$ increases, the power consumptions of Algorithm~\ref{alg:minP}, Algorithm~\ref{alg:minPSC}, the sensing-only scheme, and the HD mode are enlarged due to the hasher requirement for radar sensing.
Moreover, compared to the conventional HD mode, Algorithm~\ref{alg:minP} yields a much lower power consumption, which validates the superiority of the proposed FD scheme. Also, it is seen that Algorithm~\ref{alg:minP} and Algorithm~\ref{alg:minPSC} achieve identical performance in the considered range of parameters.

\begin{figure}[t]
\begin{center}
      \epsfxsize=7.0in\includegraphics[scale=0.6]{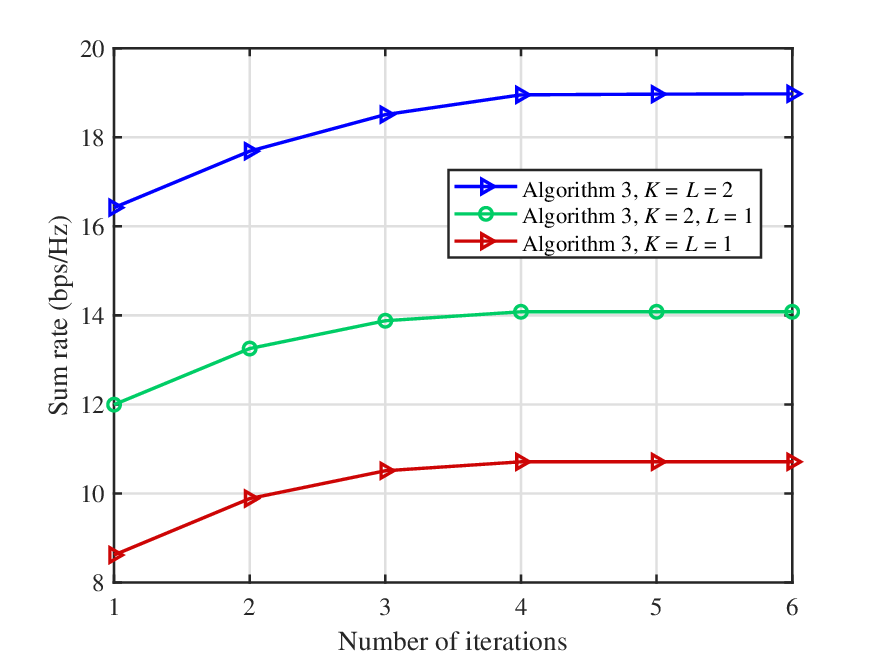}
      \caption{Convergence performance of Algorithm~3.}\label{fig:iteration_maxR}
    \end{center}
\end{figure}

\begin{figure}[t]
\begin{center}
      \epsfxsize=7.0in\includegraphics[scale=0.6]{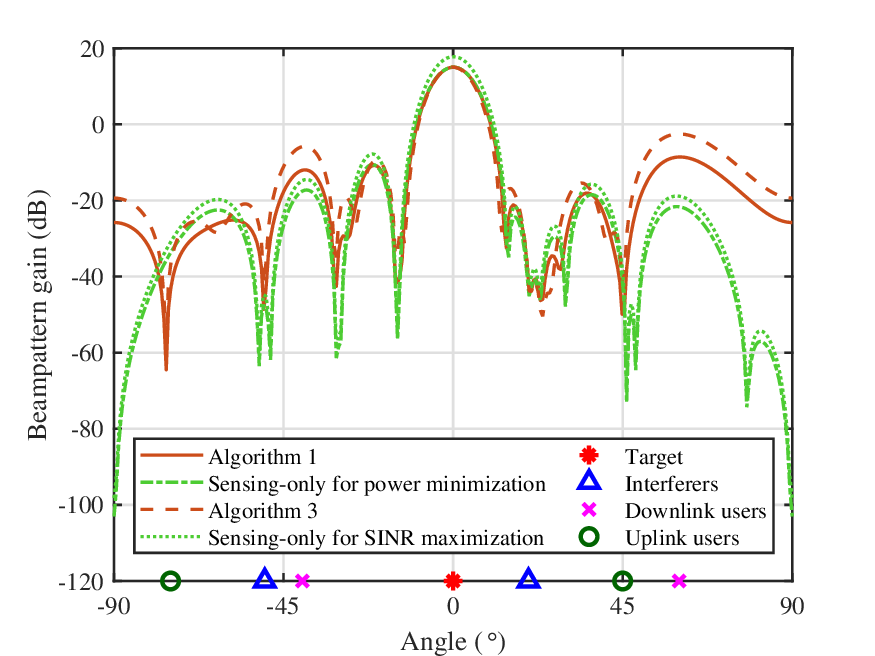}
       \caption{Beampatterns regarding radar functionality of Algorithm~1 and Algorithm~3.}\label{fig:beampattern_maxR}
    \end{center}
\end{figure}

\begin{figure}[t]
\begin{center}
      \epsfxsize=7.0in\includegraphics[scale=0.6]{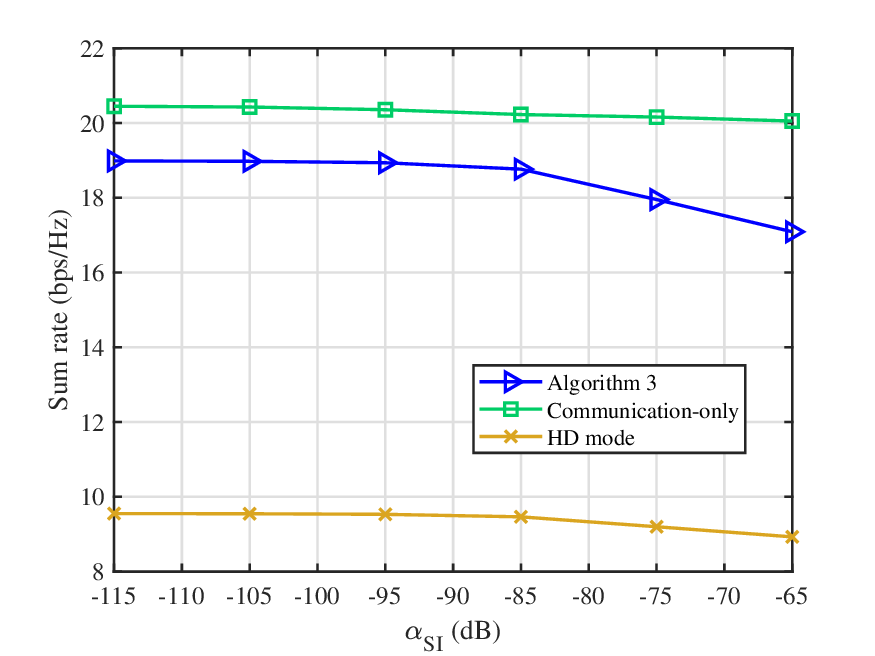}
     \caption{Multiuser sum rate versus the residual SI gain $\alpha_\text{SI}$.}\label{fig:rate_gammaSI}
    \end{center}
\end{figure}
In the following part, we evaluate the performance of Algorithm 3.
First, the convergence performance is illustrated in Fig. \ref{fig:iteration_maxR}. It can be seen that the proposed algorithm converges within a few iterations and a higher sum rate is generally obtained with the increasing number of users as the proposed scheme can exploit the inherent multiuser diversity.

The radar beampattern $p_3(\theta)$ attained by Algorithm~3 is depicted in Fig.~\ref{fig:beampattern_maxR}, where Algorithm~1 is also included for comparison.
Obviously, focusing on the curve of Algorithm~3, it is seen that the main beams are allocated to the target and to the downlink users, respectively, and meanwhile the directions of interferers and uplink users are placed with relatively deep nulls, which is consistent with the result of Fig. \ref{fig:radarpattern_of_Alg1}.
In addition, compared to the proposed algorithms, the sensing-only schemes (the green curves) have relatively lower average gains since they omit the communication requirements of users. Finally, comparing the curves of Algorithm~1 and Algorithm~3, we
observe that the average value of the beampattern gain achieved by Algorithm~3 is larger than that of Algorithm~1. This is because the objective of Algorithm~3 is to maximize the sum rate and all the available transmit power should be exhausted while Algorithm~1 aims to improve the power efficiency and full-power transmission is not always the best strategy.

Fig. \ref{fig:rate_gammaSI} demonstrates the maximum multiuser sum rate versus the residual SI gain $\alpha_\text{SI}$.
It can be observed from the figure that when $\alpha_\text{SI}$ becomes larger, the rate performances of all three schemes degrade due to the increasing power of the signal-dependent interference. This observation conforms to the results in FD communication-only systems \cite{FDcommun}.
On the other hand, compared to the communication-only scheme and the HD scheme, the achievable rate of Algorithm~3 is more sensitive to the value of $\alpha_\text{SI}$. This is because both the FD communication and the FD radar functionalities in Algorithm~3 are affected by the residual SI gain, while only the sensing functionality (the communication functionality) is affected by the SI in HD mode (in communication-only mode).
Finally, the proposed FD ISAC scheme can significantly outperform the benchmark HD mode. In particular, the achieved sum rate is nearly doubled when the SI power is relatively low.

\begin{figure}[t]
\begin{center}
      \epsfxsize=7.0in\includegraphics[scale=0.6]{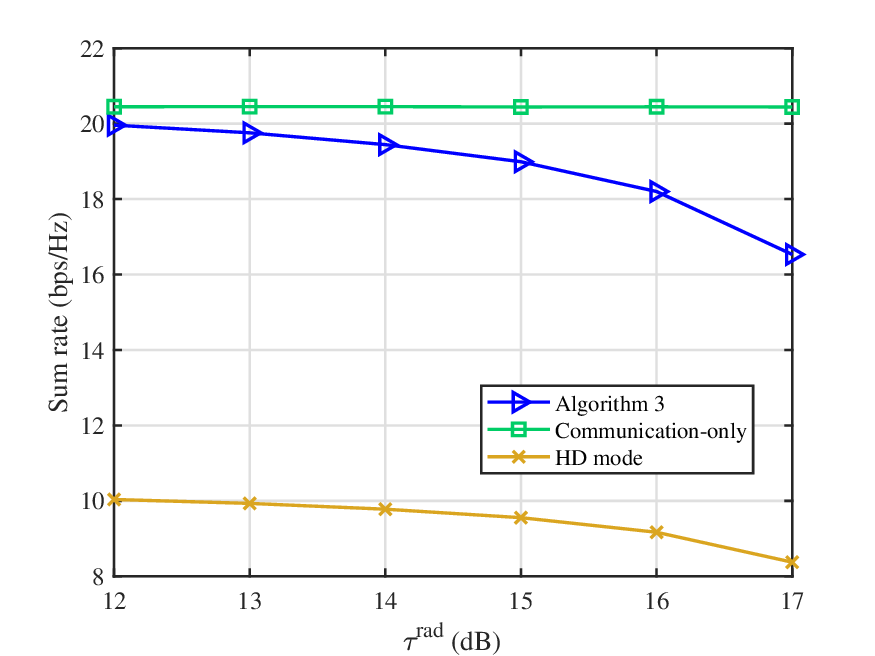}
      \caption{Multiuser sum rate versus the radar SINR threshold $\tau^\text{rad}$.}\label{fig:Rate_tau}
    \end{center}
\end{figure}

Fig. \ref{fig:Rate_tau} demonstrates the sum rate versus the required radar SINR $\tau^\text{rad}$.
We observe that when $\tau^\text{rad}$ becomes larger, the performances of Algorithm~3 and HD mode degrade. This is because with a growing $\tau^\text{rad}$, more transmit power in the ISAC system should be exploited for guaranteeing the increasingly stringent sensing requirement and meanwhile the communication rate has to be compromised, which reflects the non-trivial communication-radar trade-off in ISAC systems. Also, compared to the benchmark HD scheme, the performance improvement of the proposed FD ISAC is seen from the figure.

\begin{figure}[t]
\begin{center}
      \epsfxsize=7.0in\includegraphics[scale=0.6]{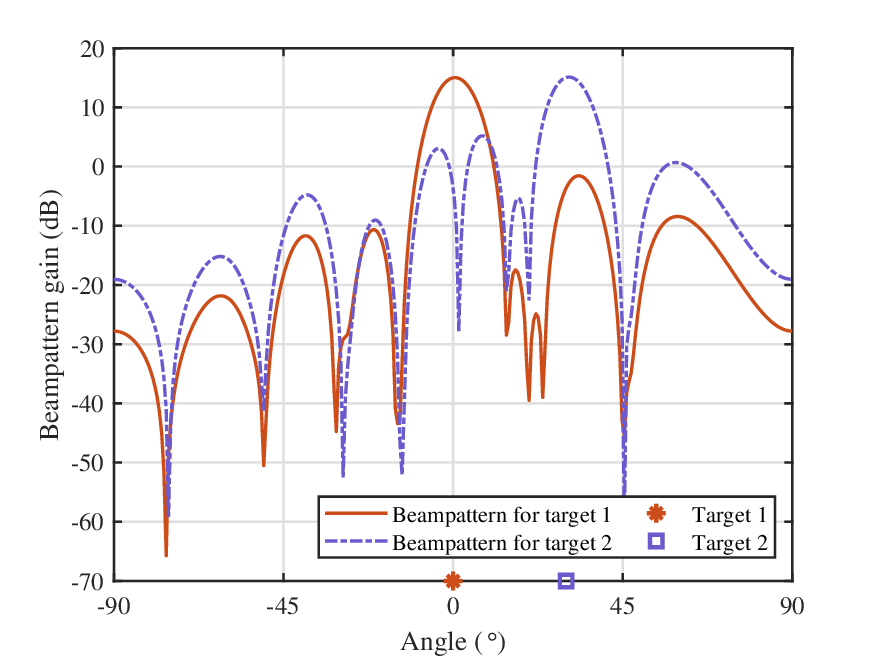}
      \caption{Radar beampattern achieved by Algorithm~1 for the multi-target scenario.}\label{fig:multitarget}
    \end{center}
\end{figure}

Finally, we provide a numerical example here to show the extension for multi-target scenarios discussed in Remark 1.
Assume $M = 2$ targets are located at the angle directions of $0^{\circ}$ and $30^{\circ}$, respectively, and the other setup parameter remains unchanged.
After solving the power minimization problem in (\ref{prob:minP}) with $M$ radar SINR requirements employing Algorithm~\ref{alg:minP}, we illustrate the achieved radar beampatterns in Fig. \ref{fig:multitarget}, where the beampattern gain for target $m$ is calculated as (\ref{def:radarpattern}) by substituting the optimized receive beamformer $\mathbf u_m^*$. It is found from the figure that the main beam of each curve is pointed to the direction of the corresponding target, which verifies the effectiveness of the proposed beamforming strategy for the multi-target detection.

\section{Conclusion}
In this paper, we investigated the joint optimization of an FD communication-based ISAC system under the criteria of transmit power minimization and sum rate maximization.
For each design problem, we first derived the optimal receive beamformers in closed-form expressions. Then, we developed an effective algorithm to optimize the BS transmit beamforming and the user transmit power based on the SCA technique. Moreover, we also considered a special case for the power minimization criterion and provided a low-cost solution, which enjoys much lower computational complexity compared to the SCA method while achieving almost identical performance.
Simulation results verified the effectiveness of the proposed algorithms and showed the tremendous advantages of our considered FD communication-based ISAC system over the previous frameworks that integrated sensing with HD communication.
The performance gains are most notable when the residual SI power is low and the sensing requirement is less restrictive.
Future extensions may include investigating the impacts of non-flat channels and imperfect CSI for the FD ISAC design.

\begin{appendices}
\section{Proof of Proposition ~\ref{prop:receiver}}\label{proof:propreceiver}
We first determine $\mathbf w_k$ for maximizing $\gamma^\text{com,UL}_k$. Based on the expression in (\ref{sinr_k}), it is found that the maximization of $\gamma^\text{com,UL}_k$ belongs to the problem of generalized Rayleigh quotient. Thereby, by invoking the results in \cite{Rayleighquotient}, we readily obtain the optimal solution to $\mathbf w_k$ in (\ref{w_k*}).
To proceed, we optimize $\mathbf u$ by rewriting the numerator of $\gamma^\text{rad}$ in (\ref{sinr_r}) as
\begin{align}
&\ |\beta_0|^2 \mathbf u^H \mathbf A(\theta_0) \mathbf Q \mathbf A(\theta_0)^H \mathbf u \nonumber\\
=&\ |\beta_0|^2 \mathbf u^H \mathbf a_r(\theta_0) \mathbf a_t^H(\theta_0) \mathbf Q \mathbf a_t(\theta_0) \mathbf a_r^H(\theta_0) \mathbf u \nonumber\\
=&\ |\beta_0|^2 \mathbf a_t^H(\theta_0) \mathbf Q \mathbf a_t(\theta_0) \mathbf u^H \mathbf a_r(\theta_0) \mathbf a_r^H(\theta_0) \mathbf u.
\end{align}
Based on the facts that $|\beta_0|^2 \mathbf a_t^H(\theta_0) \mathbf Q \mathbf a_t(\theta_0) \geq 0$ and it is not related to $\mathbf u$, we utilize the generalized Rayleigh quotient again to maximize
$
\frac{\mathbf u^H \mathbf a_r(\theta_0) \mathbf a_r^H(\theta_0) \mathbf u}
{\mathbf u^H \left( \sum_{k=1}^K p_k  \mathbf h_k \mathbf h_k^H +  \mathbf B \mathbf Q \mathbf B^H + \sigma_r^2 \mathbf I_{N_r} \right)\mathbf u}
$
and arrive at the optimal $\mathbf u^*$ in (\ref{w_r*}).

\section{Proof of Theorem ~\ref{theorem:rank1soltuion}}\label{proof:theoremrank1soltuion}
The proof follows a similar procedure in the previous work~\cite{XLiuTSP2020}. We divide the proof into two parts. {The first part is to show that $\{\{\mathbf {V}^*_l\}_{0=1}^L,\{p_k^* \}_{k=1}^K\}$ is a feasible solution to (\ref{prob:minPtildeRef}) and the second part verifies that it achieves the identical performance as $\{\{\mathbf {\widehat V}_l\}_{0=1}^L,\{\hat p_k \}_{k=1}^K\}$.}

We first prove the feasibility of $\{\{\mathbf {V}^*_l\}_{0=1}^L,\{p_k^* \}_{k=1}^K\}$.
To begin with, it is straightforwardly seen that $\mathbf {V}^*_l$ is positive semidefinite for $l=1,\cdots,L$. Moreover, given an arbitrary vector $\mathbf f \in \mathbb C^{N_t \times 1}$, it holds that
\begin{align}\label{Vhat-Vtilde}
 \mathbf f^H  \left( \mathbf {\widehat V}_l - \mathbf {v}^*_l {\mathbf {v}^*_l}^H \right) \mathbf f
=&\ \mathbf f^H \left( \mathbf {\widehat V}_l - (\mathbf g_l^H \mathbf {\widehat V}_l \mathbf g_l )^{-1} \mathbf {\widehat V}_l \mathbf g_l \mathbf g_l^H \mathbf {\widehat V}_l^H \right) \mathbf f\nonumber\\
=&\ \mathbf f^H\mathbf {\widehat V}_l\mathbf f - (\mathbf g_l^H \mathbf {\widehat V}_l \mathbf g_l )^{-1} |\mathbf f^H \mathbf {\widehat V}_l \mathbf g_l|^2 \nonumber\\
\geq&\ \mathbf f^H\mathbf {\widehat V}_l\mathbf f - (\mathbf g_l^H \mathbf {\widehat V}_l \mathbf g_l )^{-1}\mathbf f^H \mathbf {\widehat V}_l \mathbf f \mathbf g_l^H \mathbf{\widehat V}_l \mathbf g_l  \nonumber\\
=&\ 0, \ \forall l \geq 1,
\end{align}
where the inequality follows from the Cauchy-Schwarz inequality. Using (\ref{Vhat-Vtilde}), together with the fact $\mathbf {\widehat V}_0 \succeq \mathbf 0$, it can be obtained from (\ref{def:tildeV0}) that $\mathbf {V}^*_0 \succeq \mathbf 0$. Thus, we have $\{\mathbf {V}^*_l \succeq \mathbf 0 \}_{0=1}^L$.
To proceed, it follows from (\ref{def:tildeV0}) that
\begin{align}\label{equ:hat+tilde}
\sum_{l=1}^L \mathbf {\widehat V}_l + \mathbf {\widehat V}_0 = \sum_{l=1}^L \mathbf {V}^*_l + \mathbf {V}^*_0.
\end{align}
This equality, together with the unchanged values of $\{p_k \}_{k=1}^K$, implies that $\{\{\mathbf {\widehat V}_l\}_{0=1}^L,\{\hat p_k \}_{k=1}^K\}$ and $\{\{\mathbf {V}^*_l\}_{0=1}^L,\{p_k^* \}_{k=1}^K\}$ yield the same
$\mathbf{\bar Q}$, $\mathbf{\Psi}$, and $\{\mathbf{\Phi}_k\}_{k=1}^K$. Moreover,
it is verified that
\begin{align}
\mathbf g_l^H \mathbf {V}^*_l \mathbf g_l =&\ \mathbf g_l^H \mathbf {v}^*_l {\mathbf {v}^*_l}^H \mathbf g_l  \nonumber\\
=&\ (\mathbf g_l^H \mathbf {\widehat V}_l \mathbf g_l )^{-1} \mathbf g_l^H  \mathbf {\widehat V}_l \mathbf g_l \mathbf g_l^H \mathbf {\widehat V}_l^H \mathbf g_l \nonumber\\
=&\ \mathbf g_l^H \mathbf {\widehat V}_l \mathbf g_l, \ \forall l.
\end{align}
Therefore, constraints (\ref{cons:minPrad}), (\ref{cons:minPUL}), and (\ref{cons:minPDL}) also hold for $\{\{\mathbf {V}^*_l\}_{0=1}^L,\{p_k^* \}_{k=1}^K\}$. The first part is proven.

We then prove that the objective values of $\{\{\mathbf {V}^*_l\}_{0=1}^L,\{p_k^* \}_{k=1}^K\}$ and $\{\{\mathbf {\widehat V}_l\}_{0=1}^L,\{\hat p_k \}_{k=1}^K\}$ are identical. {Applying (\ref{def:tildeV0}), we immediately have
$ \sum_{l=0}^L \text{Tr}(\mathbf {\widehat V}_l) + \sum_{k=1}^K \hat p_k= \sum_{l=0}^L \text{Tr}(\mathbf {V}^*_l)+ \sum_{k=1}^K p_k^*$. The second part of proof is completed.}

\section{Proof of Theorem ~\ref{theorem:V0rank}}\label{proof:theoremV0rank}
We accomplish the proof by analyzing the KKT conditions of the problem in (\ref{prob:minP_SCSDP}). Since (\ref{prob:minP_SCSDP}) is a convex SDP and the Slater's condition holds, the duality gap is zero and the KKT conditions are sufficient and necessary for guaranteeing the optimality \cite{cvx}.
Let $\mu \geq 0$, $\{ \lambda_k \geq 0 \}_{k=1}^K$, and a positive semidefinite matrix $\mathbf Z \succeq \mathbf 0$ denote the Lagrange multipliers associated with the radar SINR constraint, the uplink communication SINR constraints, and the semidefinition constraint $\mathbf V_0 \succeq \mathbf 0$, respectively. Thus, the partial Lagrangian function of (\ref{prob:minP_SCSDP}) is given by
\begin{align}
&\ \mathcal L( \mathbf V_0,\{ p_k \}_{k=1}^K, \mu,\{ \lambda_k \}_{k=1}^K, \mathbf Z) \nonumber \\
=&\ \text{Tr}(\mathbf V_0) \!+\! \sum_{k=1}^K p_k \!+\! \mu (\mathbf{\tilde f}^H \mathbf V_0 \mathbf{\tilde f} \!+\! \sum_k \tilde g_k p_k + \tilde h \!-\! \frac{1}{\tau^\text{rad}} \mathbf{\tilde e}^H \mathbf V_0 \mathbf{\tilde e}) \nonumber \\
&\ +\! \sum_{k=1}^K \lambda_k (\mathbf{\tilde b}_k^H \mathbf V_0 \mathbf{\tilde b}_k \!+\! \sum_{k'\neq k} \tilde{c}_{k,k'} p_{k'} \!+\! \tilde d_k \!-\! \frac{1}{\tau^\text{com,UL}_k} \tilde a_kp_k )  \nonumber \\
&\ - \text{Tr}(\mathbf Z \mathbf V_0).
\end{align}
Given the Lagrangian function, we further obtain the dual function of problem (\ref{prob:minP_SCSDP}) by
\begin{align}\label{dualfunction}
&\ g(\mu,\{ \lambda_k \}_{k=1}^K, \mathbf Z)  \nonumber \\
=&\ \mathop \text{inf} \limits_{ \mathbf V_0, \{ p_k\geq0 \}_{k=1}^K} \ \mathcal L( \mathbf V_0,\{ p_k\}_{k=1}^K,\mu,\{ \lambda_k \}_{k=1}^K, \mathbf Z) \nonumber\\
 =&\
\mathop \text{inf} \limits_{ \mathbf V_0,\{ p_k\geq0 \}_{k=1}^K} \ \text{Tr}(\mathbf B \mathbf V_0) + f(\{ p_k\}_{k=1}^K),
\end{align}
where
$\mathbf B = \mathbf I_{N_t} + \mu \mathbf{\tilde f} \mathbf{\tilde f}^H + \sum_{k=1}^K \lambda_k \mathbf b_k \mathbf b_k^H -  \frac{\mu}{\tau^\text{rad}} \mathbf{\tilde e} \mathbf{\tilde e}^H  - \mathbf Z$ {and $f(\{ p_k\}_{k=1}^K)$ contains the remaining terms related to $\{ p_k\}_{k=1}^K$.}
In order to guarantee a bounded dual optimal value, it follows from (\ref{dualfunction}) that $\mathbf B = \mathbf 0$, which means that
\begin{align}\label{Z}
\mathbf Z = \mathbf I_{N_t} + \mu \mathbf{\tilde f} \mathbf{\tilde f}^H + \sum_{k=1}^K \lambda_k \mathbf b_k \mathbf b_k^H -  \frac{\mu}{\tau^\text{rad}} \mathbf{\tilde e} \mathbf{\tilde e}^H.
\end{align}
Furthermore, due to the non-negativeness of $ \mu$ and $\{ \lambda_k \}_{k=1}^K$, we can infer from (\ref{Z}) that
$\text{rank} (\mathbf Z) \geq N_t -1$.
On the other hand, we list the related KKT conditions of problem (\ref{prob:minP_SCSDP}) for the proof as follows
\begin{align}
 \frac{1}{\tau^\text{rad}} \mathbf{\tilde e}^H \mathbf V_0^* \mathbf{\tilde e} \geq &\ \mathbf{\tilde f}^H \mathbf V_0^* \mathbf{\tilde f} + \sum_k \tilde g_k p_k^* + \tilde h, \label{KKT1}\\
\mathbf Z^* \mathbf V_0^* = &\ \mathbf 0 .\label{KKT2}
\end{align}
Combining $\text{rank} (\mathbf Z) \geq N_t -1$ and (\ref{KKT2}), we have $\text{rank} (\mathbf V_0^*) \leq 1$. Moreover, condition (\ref{KKT1}) holds only when $\text{rank} (\mathbf V_0^*) \neq 0$ since $\tilde h >0$. Thereby, we conclude that
$\text{rank} (\mathbf V_0^*) = 1.$

\section{Proof of Proposition~\ref{prop:minPSOCP}}\label{proof:propminPSOCP}
It follows from Theorem~\ref{theorem:V0rank} that we can express $\mathbf V_0$ by $\mathbf V_0 = \mathbf v_0 \mathbf v_0^H$ without loss of optimality, where $\mathbf v_0 \in \mathbb C^{N_t \times 1}$ represents the radar beamformer.
Denoting $q_k = \sqrt{p_k} \geq 0,\ \forall k,$ problem (\ref{prob:minP_SCSDP}) is recast as
\begin{align}\label{prob:minP_sensingonlyv0}
\mathop \text{minimize} \limits_{\mathbf v_0, \{q_k\}_{k=1}^K} \quad
& \|\mathbf v_0 \|^2 + \sum_{k=1}^K q_k^2 \nonumber \\
\text{subject to}\quad
& \frac{ |\mathbf {\tilde e}^H\mathbf v_0 |^2}
{|\mathbf {\tilde f}^H \mathbf v_0|^2 + \sum_{k} {\tilde g}_kq_k^2 + {\tilde h}} \geq \tau^\text{rad}, \nonumber \\
& \frac{ {\tilde a}_kq_k^2 }{ |\mathbf {\tilde b}_k^H \mathbf v_0|^2 + \sum_{k'\neq k}{\tilde c}_{k,k'}q_{k'}^2 + {\tilde d}_k } \geq \tau_k^\text{com,UL},  \ \forall k.
\end{align}
For (\ref{prob:minP_sensingonlyv0}), it can be verified that rotating the optimal $\mathbf v_0$ with
an arbitrary phase scaling does not destroy the optimality. Therefore, focusing on the radar SINR in the first constraint of (\ref{prob:minP_sensingonlyv0}), it is without loss of optimality to further restrict
$\mathcal{I}\{\mathbf {\tilde e}^H \mathbf v_0\} = 0$.
Then, we can take the square root of $|\mathbf {\tilde e}^H \mathbf v_0 |^2$ and transform the first constraint in (\ref{prob:minP_sensingonlyv0}) to
$ \mathbf {\tilde e}^H \mathbf v_0 \geq \left\| \boldsymbol \varpi \right\|$, where $\boldsymbol \varpi \triangleq \sqrt{\tau^\text{rad}}\left[ \mathbf {\tilde f}^H \mathbf v_0, \sqrt{{\tilde g}_1} q_1, \cdots, \sqrt{{\tilde g}_K} q_K, \sqrt{{\tilde h}} \right]^T $.
It can be further rewritten as the following SOC
$
\left[ \begin{matrix} \mathbf {\tilde e}^H \mathbf v_0 \\ \boldsymbol \varpi \end{matrix} \right] \succeq_C \mathbf 0.
$
To proceed, performing the similar operations for the communication SINR constraints in (\ref{prob:minP_sensingonlyv0}) yields the second set of SOCs in (\ref{prob:minP_sensingonlySOCP}).
Finally, by introducing an auxiliary variable $t > 0$, the minimization of $\| \mathbf v_0 \|^2 + \sum_{k=1}^K q_k^2$ is equivalent to minimize $t$ with an additional constraint $t \geq \| \mathbf v_0 \|^2 + \sum_{k=1}^K q_k^2$. Then,
denoting $t_0 = \sqrt{t}$, (\ref{prob:minP_sensingonlyv0}) is transformed into the SOCP given in (\ref{prob:minP_sensingonlySOCP}).

\section{Details of The TDD Benchmark Scheme}\label{appendix:HD}
Assume that the uplink and downlink time slots have the same duration.
In the downlink slot, the BS transmits an ISAC signal $\mathbf x$ and receives the radar echo signal adopting a linear beamformer $\mathbf u$. The involved optimization problems of transmit power minimization and sum rate maximization take the form:
\begin{align}
(\mathcal P_\text{minP}^\text{HD,DL}): \mathop \text{minimize} \limits_{\{\mathbf v_l \}_{l=1}^L, \mathbf V_0 \succeq \mathbf 0, \mathbf u}
& \sum_{l=1}^L \|\mathbf v_l \|^2 + \text{Tr}(\mathbf V_0) \nonumber \\
\text{subject to}\quad
& \gamma^\text{rad,DL}_\text{HD} \geq \tau^\text{rad}, \nonumber \\
& \gamma^\text{com,DL}_l \geq \bar \tau^\text{com,DL}_l,\ \forall l.\\
(\mathcal P_\text{maxR}^\text{HD,DL}):\mathop \text{maximize} \limits_{\{\mathbf v_l \}_{l=1}^L, \mathbf V_0 \succeq \mathbf 0, \mathbf u} \
& \sum_{l=1}^L \log_2(1 + \gamma_l^\text{com,DL}) \nonumber \\
\text{subject to}\quad
& \gamma^\text{rad,DL}_\text{HD} \geq \tau^\text{rad}, \nonumber \\
& \sum_{l=1}^L \| \mathbf v_l \|^2 + \text{Tr}(\mathbf V_0) \leq P_\text{max},
\end{align}
respectively. Here, $\gamma^\text{rad,DL}_\text{HD} = \frac{|\beta_0|^2 \mathbf u^H \mathbf A(\theta_0) \mathbf Q \mathbf A(\theta_0)^H \mathbf u}{\mathbf u^H \left(\mathbf B \mathbf Q \mathbf B^H + \sigma^2_r \mathbf I_{N_r} \right)\mathbf u}$ represents the radar SINR without the interference from the uplink transmission due to the HD mode and $ \bar \tau^\text{com,DL}_l \triangleq (1+\tau^\text{com,DL}_l)^2-1$ ensures that the minimum average data rate of downlink user $l$ achieved in the HD scheme equals to that of the FD case.
In the uplink slot, the BS simultaneously receives the communication signal from $K$ uplink users and transmits $\mathbf s_0$ for downlink sensing, as discussed in Section III-C. The joint optimization problems are written as
\begin{align}
(\mathcal P_\text{minP}^\text{HD,UL}):\mathop \text{minimize} \limits_{ \mathbf V_0 \succeq \mathbf 0,\atop \{\mathbf w_k,p_k \geq0 \}_{k=1}^K} \quad
& \text{Tr}(\mathbf V_0) + \sum_{k=1}^K p_k \nonumber \\
\text{subject to}\quad \
& \gamma^\text{rad} \geq \tau^\text{rad}, \nonumber \\
& \gamma^\text{com,UL}_k \geq \bar \tau^\text{com,UL}_k,\ \forall k. \\
(\mathcal P_\text{maxR}^\text{HD,UL}):\mathop \text{maximize} \limits_{ \mathbf V_0 \succeq \mathbf 0,\atop \{\mathbf w_k,p_k \geq0 \}_{k=1}^K} \quad
& \sum_{k=1}^K \log_2(1 + \gamma_k^\text{com,UL}) \nonumber \\
\text{subject to}\quad\
& \gamma^\text{rad} \geq \tau^\text{rad},\nonumber \\
& \text{Tr}(\mathbf V_0) \leq P_\text{max}, \nonumber \\
& p_k \leq P_k,\ \forall k,
\end{align}
respectively, where $ \bar \tau^\text{com,UL}_k \triangleq (1+\tau^\text{com,UL}_k)^2-1$ guarantees that the minimum average rates of uplink communication in both FD and HD cases are identical.
Denote the optimized objective values of these four problems as $P^\text{DL}$, $R^\text{DL}$, $P^\text{UL}$, and $R^\text{UL}$, respectively. We finally obtain the average power consumption and achievable rate of this TDD system by $\frac{1}{2} (P^\text{DL} + P^\text{UL})$ and $\frac{1}{2} (R^\text{DL} + R^\text{UL})$, respectively.
\end{appendices}


\begin{thebibliography}{99}
\bibitem{ICASSP}
Z. He \textit{et al.}, ``Integrated sensing and full-duplex communication: Joint transceiver beamforming and power allocation,'' in \emph{Proc. IEEE Int. Conf. Acoust., Speech Signal Process. (ICASSP)}, Rhodes Island, Greece, Jun.~2023.

\bibitem{W.XuJSTSP2022}
W. Xu \textit{et al.}, ``Edge learning for B5G networks with distributed signal processing: Semantic communication, edge computing, and wireless sensing,'' \emph{IEEE J. Sel. Topics Signal Process.}, vol. 17, no. 1, pp. 9--39, Jan. 2023.

\bibitem{W.ShiWCM2022}
W. Shi, W. Xu, X. You, C. Zhao, and K. Wei, ``Intelligent reflection enabling technologies for integrated and green Internet-of-Everything beyond 5G: Communication, sensing, and security,'' \emph{IEEE Wireless Commun.}, early access, May 09, 2022, doi: 10.1109/MWC.018.2100717.

\bibitem{overviewJ.Zhang}
J. A. Zhang \textit{et al.}, ``An overview of signal processing techniques for joint communication and radar sensing,'' \emph{IEEE J. Sel. Topics Signal Process.}, vol. 15, no. 6, pp. 1295--1315, Nov. 2021.

\bibitem{J.Zhang2022CSTEnabling}
J. A. Zhang \textit{et al.}, ``Enabling joint communication and radar sensing in mobile networks---A survey,'' \emph{IEEE Commun. Surveys Tuts.,} vol. 24, no.~1, pp. 306--345, 1st Quart. 2022.

\bibitem{overviewF.Liu}
F. Liu \textit{et al.}, ``Integrated sensing and communications: Towards dual-functional wireless networks for 6G and beyond,'' \emph{IEEE J. Sel. Areas Commun.}, vol. 40, no. 6, pp. 1728--1767, Jun. 2022.

\bibitem{P.StoicaMIMOradar1}
J. Li and P. Stoica, ``MIMO radar with colocated antennas,'' \emph{IEEE Signal Process. Mag.}, vol. 24, no. 5, pp. 106--114, Sep. 2007.

\bibitem{P.StoicaTSP2007}
P. Stoica, J. Li, and Y. Xie, ``On probing signal design for MIMO radar,'' \emph{IEEE Trans. Signal Process.}, vol. 55, no. 8, pp. 4151--4161, Aug. 2007.

\bibitem{F.LiuTWC2018}
F. Liu \textit{et al.}, ``MU-MIMO communications with MIMO radar: From coexistence to joint transmission,'' \emph{IEEE Trans. Wireless Commun.}, vol.~17, no.~4, pp. 2755--2770, Apr. 2018.

\bibitem{XLiuTSP2020}
X. Liu, T. Huang, N. Shlezinger, Y. Liu, J. Zhou, and Y. C. Eldar, ``Joint transmit beamforming for multiuser MIMO communications and MIMO radar,'' \emph{IEEE Trans. Signal Process.}, vol. 68, pp. 3929--3944, Jun. 2020.

\bibitem{HHuaArxiv}
H. Hua, J. Xu, and T. X. Han, ``Optimal transmit beamforming for integrated sensing and communication,'' \emph{IEEE Trans. Veh. Technol.}, early access, Mar. 29, 2023, doi: 10.1109/TVT.2023.3262513.

\bibitem{Z.LyuArxiv}
Z. Lyu, G. Zhu, and J. Xu, ``Joint maneuver and beamforming design for UAV-enabled integrated sensing and communication,'' \emph{IEEE Trans. Wireless Commun.}, vol. 22, no. 4, pp. 2424--2440, Apr. 2023.

\bibitem{Z.HeWCL2022}
Z. He, W. Xu, H. Shen, Y. Huang, and H. Xiao, ``Energy efficient beamforming optimization for integrated sensing and communication,'' \emph{IEEE Wireless Commun. Lett.}, vol. 11, no. 7, pp. 1374--1378, Jul. 2022.

\bibitem{Liu2022CRB}
F. Liu, Y.-F. Liu, A. Li, C. Masouros, and Y. C. Eldar, ``Cram\'er-rao bound optimization for joint radar-communication beamforming,'' \emph{IEEE Trans. Signal Process.}, vol. 70, pp. 240--253, Jan. 2022.

\bibitem{J.Pritzker2022Arxiv}
J. Pritzker, J. Ward, and Y. C. Eldar, ``Transmit precoder design approaches for dual-function radar-communication systems,'' 2022. Available: https://arxiv.org/pdf/2203.09571.pdf

\bibitem{L.Chen2022JSAC}
L. Chen, Z. Wang, Y. Du, Y. Chen, and F. Richard Yu, ``Generalized transceiver beamforming for DFRC with MIMO radar and MU-MIMO communication,'' \emph{IEEE J. Sel. Areas Commun.}, vol. 40, no. 6, pp. 1795--1808, Jun. 2022.

\bibitem{C.Tsinos2021JSTSP}
C. G. Tsinos, A. Arora, S. Chatzinotas, and B. Ottersten, ``Joint transmit waveform and receive filter design for dual-function radar-communication systems,'' \emph{IEEE J. Sel. Topics Signal Process.}, vol. 15, no. 6, pp. 1378--1392, Nov. 2021.


\bibitem{M.Temiz2021TCCN}
M. Temiz, E. Alsusa, and M. W. Baidas, ``A dual-function massive MIMO uplink OFDM communication and radar architecture,'' \emph{IEEE Trans. Cogn. Commun. Netw.}, vol. 8, no. 2, pp. 750--762, Jun. 2022.

\bibitem{X.Wang2022CL}
X. Wang, Z. Fei, J. A. Zhang, and J. Huang, ``Sensing-assisted secure uplink communications with full-duplex base station,''
\emph{IEEE Commun. Lett.}, vol. 26, no. 2, pp. 249--253, Feb. 2022.

%
%


\bibitem{FDISAC}
C. B. Barneto \textit{et al.}, ``Full-duplex OFDM radar with LTE and 5G NR waveforms: Challenges, solutions, and
measurements,'' \emph{IEEE Trans. Microw. Theory Techn.}, vol. 67, no. 10, pp. 4042--4054, Oct. 2019.

\bibitem{C.B.BarnetoWCM2021Full}
C. B. Barneto, S. D. Liyanaarachchi, M. Heino, T. Riihonen, and M. Valkama, ``Full duplex radio/radar technology: The enabler for advanced joint communication and sensing,'' \emph{IEEE Wireless Commun.}, vol.~28, no.~1, pp.~82--88, Feb. 2021.

\bibitem{FDSI1}
A. Sabharwal, P. Schniter, D. Guo, D. W. Bliss, S. Rangarajan, and R. Wichman, ``In-band full-duplex wireless: Challenges and opportunities,'' \emph{IEEE J. Sel. Areas Commun.}, vol. 32, no. 9, pp. 1637--1652, Sep. 2014.

\bibitem{FDSI5}
K. E. Kolodziej, B. T. Perry, and J. S. Herd, ``In-band full-duplex technology: Techniques and systems survey,'' \emph{IEEE
Trans. Microw. Theory Techn.}, vol. 67, no. 7, pp. 3025--3041, Jul. 2019.

\bibitem{RadarBook}
M. A. Richards, J. Scheer, W. A. Holm, and W. L. Melvin, \emph{Principles of Modern Radar: Basic Principles.} Chennai, India: SciTech Publishing, 2010.






\bibitem{C.Y.ChenTSP2009MIMOradar}
C.-Y. Chen and P. P. Vaidyanathan, ``MIMO radar waveform optimization with prior information of the extended target and clutter,'' \emph{IEEE Trans. Signal Process.}, vol. 57, no. 9, pp. 3533--3544, Sep. 2009.

\bibitem{Wu2018Transmit}
L. Wu, P. Babu, and D. P. Palomar, ``Transmit waveform/receive filter design for MIMO radar with multiple waveform constraints,''
\emph{IEEE Trans. Signal Process.}, vol. 66, no. 6, pp. 1526--1540, Mar. 2018.

\bibitem{radarCM}
G. Cui, H. Li, and M. Rangaswamy, ``MIMO radar waveform design with constant modulus and similarity constraints,'' \emph{IEEE Trans. Signal Process.}, vol. 62, no. 2, pp. 343--353, Jan. 2014.

\bibitem{Multitarget}
X. Yu, K. Alhujaili, G. Cui, and V. Monga, ``MIMO radar waveform design in the presence of multiple targets and practical constraints,'' \emph{IEEE Trans. Signal Process.}, vol. 68, pp. 1974--1989, Apr. 2020.




\bibitem{GaussianR}
Z.-Q. Luo, W.-K. Ma, A. M.-C. So, Y. Ye, and S. Zhang, ``Semidefinite relaxation of quadratic optimization problems,'' \emph{IEEE Signal Process. Mag.}, vol. 27, no. 3, pp. 20--34, May. 2010.

\bibitem{cvx}
S. Boyd and L. Vandenberghe, \textit{Convex Optimization}. Cambridge, U.K.:
Cambridge Univ. Press, 2004.

\bibitem{CCCP}
T. Lipp and S. Boyd, ``Variations and extension of the convex-concave
procedure,'' \textit{Optim. Eng.}, vol. 17, no. 2, pp. 263--278, Jun. 2016.

\bibitem{Derivatives}
A. Hjoungnes, \textit{Complex-Valued Matrix Derivatives: With Applications in
Signal Processing and Communications}. Cambridge, U.K.: Cambridge
Univ. Press, 2011.


\bibitem{CVXtool}
M. Grant and S. Boyd. (Dec. 2018). \textit{CVX: MATLAB Software for Disciplined
Convex Programming.} [Online]. Available: http://cvxr.com/cvx/

\bibitem{SCAconvergence}
A. Beck, A. Ben-Tal, and L. Tetruashvili, ``A sequential parametric convex approximation method with applications to nonconvex truss topology design problems,'' \emph{J. Global Optim.}, vol. 47, no. 1, pp. 29--51, May 2010.

\bibitem{complexity}
K.-Y. Wang, A. M.-C. So, T.-H. Chang, W.-K. Ma, and C.-Y. Chi, ``Outage constrained robust transmit optimization for multiuser MISO downlinks: Tractable approximations by conic optimization,'' \emph{IEEE Trans. Signal Process.}, vol. 62, no. 21, pp. 5690--5705, Nov. 2014.

\bibitem{A.Wiesel2006Linear}
A. Wiesel, Y. C. Eldar, and S. Shamai (Shitz), ``Linear precoding via conic optimization for fixed MIMO receivers,'' \emph{IEEE Trans. Signal Process.}, vol. 54, no. 1, pp. 161--176, Jan. 2006.

\bibitem{LuoJSTSP2008Dynamic}
Z.-Q. Luo and S. Zhang, ``Dynamic spectrum management: Complexity and duality,'' \emph{IEEE J. Sel. Topics Signal Process.,} vol. 2, no. 1, pp. 57--73, Feb. 2008.

\bibitem{C.XingTSP2020Newpoint}
C. Xing, Y. Jing, S. Wang, S. Ma, and H. V. Poor, ``New viewpoint and algorithms for water-filling solutions in wireless communications,'' \emph{IEEE Trans. Signal Process.}, vol. 68, pp. 1618--1634, Feb. 2020.



%

\bibitem{O.E.Ayach2014TWC}
O. E. Ayach, S. Rajagopal, S. Abu-Surra, Z. Pi, and R. W. Heath, ``Spatially sparse precoding in millimeter wave MIMO systems,'' \emph{IEEE Trans. Wireless Commun.}, vol. 13, no. 3, pp. 1499--1513, Mar. 2014.

\bibitem{X.Chen2018Access}
X. Chen, S. Zhang, and Q. Li, ``A review of mutual coupling in MIMO systems,'' \emph{IEEE Access}, vol. 6, pp. 24706--24719, 2018.

\bibitem{FDcommun}
D. Nguyen, L.-N. Tran, P. Pirinen, and M. Latva-Aho, ``On the spectral efficiency of full-duplex small cell wireless systems,'' \emph{IEEE Trans. Wireless Commun.}, vol. 13, no. 9, pp. 4896--4910, Sep. 2014.



\bibitem{A.Maiod2008Code}
A. De Maio \textit{et al.}, ``Code design to optimize radar detection performance under accuracy and similarity constraints,'' \emph{IEEE Trans. Signal Process.}, vol. 56, no. 11, pp. 5618--5629, Nov. 2008.

\bibitem{Rayleighquotient}
G. H. Golub and C. F. Van Loan, \textit{Matrix Computations}, 3rd ed.
Baltimore, MD, USA: Johns Hopkins Univ. Press, 1996.
\end{thebibliography}
\end{document}